\documentclass[twocolumn,showpacs,pra,floatfix]{revtex4}

\usepackage{graphicx}
\usepackage{bm}
\usepackage{psfrag}
\usepackage{amsmath}
\usepackage{amssymb}
\usepackage{latexsym}
\usepackage{exscale}

\newcommand{\vect}[1]{\bm{#1}}
\newcommand{\ten}[1]{\mbox{\textbf{
{\textsf{#1}}}}}

\newcommand{\veczero}{\mbox{\textbf{\textit{0}}}}
\newcommand{\tenszero}{\mbox{\textbf{\textsf{0}}}}
\newcommand{\sprod}{\!\cdot\!}

\newcommand{\trace}{\operatorname{tr}}
\newcommand{\trans}{\mathsf{T}}
\newcommand{\dif}{\mathrm{d}}
\newcommand{\mi}{\mathrm{i}}
\newcommand{\me}{\mathrm{e}}

\begin{document}

\title{Surface-induced heating of cold polar molecules}

\author{Stefan Yoshi Buhmann}
\author{M. R. Tarbutt}
\author{Stefan Scheel}
\author{E. A. Hinds}
\affiliation{Quantum Optics and Laser Science, Blackett Laboratory,
Imperial College London, Prince Consort Road,
London SW7 2AZ, United Kingdom}

\date{\today}

\begin{abstract}
We study the rotational and vibrational heating of diatomic molecules
placed near a surface at finite temperature on the basis of
macroscopic quantum electrodynamics. The internal molecular evolution
is governed by transition rates that depend on both temperature and
position. Analytical and numerical methods are used to investigate the
heating of several relevant molecules near various surfaces. We
determine the critical distances at which the surface itself becomes
the dominant source of heating and we investigate the transition
between the long-range and short-range behaviour of the heating rates.
A simple formula is presented that can be used to estimate the
surface-induced heating rates of other molecules of interest. We also
consider how the heating depends on the thickness and composition of
the surface.
\end{abstract}

\maketitle


\section{Introduction}
\label{Sec1}
A number of techniques have recently been developed to cool polar
molecules to low temperatures and to trap them for a second or longer.
Using the switched electric field gradients of a Stark decelerator
\cite{Bethlem(1)99}, polar molecules formed in a supersonic expansion
have been decelerated to rest and then stored in electrostatic,
magnetic or electrodynamic traps \cite{Bethlem(1)00, Meerakker(1)05,
Sawyer(1)07, Veldhoven(1)05}. An electrostatic trap has been
continuously loaded by filtering out the slowest fraction of the
molecules present in an effusive beam \cite{Rieger(1)05}. Polar
molecules have also been cooled in a buffer gas of cold helium and
then confined in a magnetic trap \cite{Weinstein(1)98}. Extremely cold
polar molecules such as RbCs can be produced by the photoassociation
of two species of ultracold atoms, followed by laser-stimulated state
transfer \cite{Sage(1)05}. In all cases, the resulting molecules are
typically far colder than their environments, and they may be heated
by the absorption of blackbody radiation from that environment. Unlike
atoms, the polar molecules can be rotationally or vibrationally
excited by their interaction with this blackbody radiation, and in
many cases this can severely limit the trapping lifetime of the
molecules. Indeed, the blackbody heating rate for trapped OH and OD
has already been measured experimentally and found to limit the
trapping lifetime to just a few seconds when the environment is at
room temperature \cite{Hoekstra(1)07}. Calculations of the free-space
heating rates for several polar molecules have already been presented
\cite{0772}.

In most experiments so far, the cold polar molecules have been
confined in macroscopic traps, with trap surfaces typically several mm
from the molecules themselves. There is now a great deal of interest
in confining and manipulating these molecules much closer to surfaces,
so as to build a `molecule chip' technology analogous to that for
atoms \cite{Fortagh(1)07}. Fast-moving molecules have already been
trapped in travelling potential wells formed approximately 25\,$\mu$m
above a microstructured surface \cite{Meek(1)08}. This same structure
can be used to decelerate the molecules to rest so that they can be
trapped above the surface of the chip. Chip-based microtraps have been
designed, along with schemes to interface the molecules with
solid-state devices which could be used to cool, detect and control
them coherently \cite{Andre(1)06}. Strong coupling to a
superconducting stripline cavity is possible when the molecules are
just a few microns from the surface, and then the molecules can be the
long-lived quantum memory of a hybrid quantum information processor
\cite{Rabl(1)06}. Recent developments also herald the prospect of
integrated molecule detectors, based on optical microcavities
\cite{Trupke(1)07} or ultrathin optical fibers \cite{Warken(1)07}.

These advances raise the question of the heating rates in the close
vicinity of a surface. The influence of such a non-trivial environment
on the internal atomic dynamics is commonly known as the Purcell
effect \cite{Purcell46}. Early theoretical studies were devoted to the
zero temperature case where the evolution is governed by spontaneous
decay. As shown by linear response theory, the decay rate for an
arbitrary environment can be given in terms of the classical Green
tensor for the respective geometry \cite{Agarwal75}. Alternative
approaches have been developed on the basis of classical
electrodynamics \cite{0335,0336} and microscopic models \cite{Yeung96}
and have been applied to the case of an atom near a single surface or
between two surfaces. Results for an arbitrary environment of electric
\cite{Scheel99} and magneto-electric bodies \cite{0002}, including
local-field effects \cite{0489,0739}, have also been obtained on the
basis of macroscopic quantum electrodynamics (QED) and have been used
to study atoms in bulk material \cite{Scheel99}, outside \cite{Ho01}
or inside a microsphere \cite{0489}, inside a spherical cavity
\cite{Ho00} and even in the presence of left-handed meta-materials
\cite{0002,0737,0828}. The linear-response approach has been
generalised to finite temperatures \cite{0041} where the internal
dynamics is no longer governed by spontaneous decay alone, but
stimulated emission and absorption of thermal photons also contribute.
The respective environment-dependent transition rates can again be
expressed in terms of the classical Green tensor; in addition, the
thermal photon number comes into play. Ground-state heating rates of
spinless atoms have been predicted to be very small near surfaces
\cite{Henkel99}, in contrast to the case of atoms with spin which have
been investigated for planar surfaces, \cite{Henkel99,Henkel99b} wires
\cite{0191} and carbon nanotubes \cite{Fermani07}.

In this paper, we calculate heating rates for a number of polar
molecules currently favoured by experimenters. On the basis of
macroscopic QED (presented in Sec.~\ref{Sec2}), we solve the internal
molecular dynamics to obtain transition rates of a molecule in an
arbitrary uniform-temperature environment (Sec.~\ref{Sec3}). In
Sec.~\ref{Sec4}, the results are first used to calculate the rates in
free space, and then as a function of distance from the surface of
some common metals and dielectrics, as well as some unusual
meta-materials.


\section{Macroscopic quantum electrodynamics at finite temperature}
\label{Sec2}

Consider a molecule (or an atom) that is placed within an arbitrary
environment of magneto-electric bodies. The coupled dynamics of the
molecule and the body-assisted electromagnetic field can be described
by the Hamiltonian \cite{0002,0696}
\begin{equation}
\label{eq2.1}
\hat{H}=\hat{H}_A+\hat{H}_F+\hat{H}_{AF},
\end{equation}
where
\begin{equation}
\label{eq2.2}
\hat{H}_A=\sum_n E_n |n\rangle\langle n|
\end{equation}
($E_n$, molecular eigenenergies; $|n\rangle$, molecular eigenstates)
is the Hamiltonian of the molecule,
\begin{equation}
\label{eq2.3}
\hat{H}_F
 =\sum_{\lambda=e,m}\int\dif^3r \int_0^\infty
 \dif\omega\,\hbar\omega\,
 \hat{\vect{f}}_\lambda^\dagger(\vect{r},\omega)
 \sprod\hat{\vect{f}}_\lambda(\vect{r},\omega),
\end{equation}
is the Hamiltonian of the electromagnetic field (including the
internal charges present in the bodies) expressed in terms of the
bosonic variables
\begin{gather}
\label{eq2.4}
\Bigl[\hat{f}_{\lambda i}(\vect{r},\omega),
 \hat{f}_{\lambda'j}(\vect{r}',\omega')\Bigr]=0
 =\Bigl[\hat{f}_{\lambda i}^\dagger(\vect{r},\omega),
 \hat{f}_{\lambda'j}^\dagger(\vect{r}',\omega')\Bigr],\\
\label{eq2.5}
\Bigl[\hat{f}_{\lambda i}(\vect{r},\omega),
 \hat{f}_{\lambda'j}^\dagger(\vect{r}',\omega')\Bigr]
 =\delta_{\lambda\lambda'}
 \delta_{ij}\delta(\vect{r}-\vect{r}')\delta(\omega-\omega'),
\end{gather}
(note that $\hat{\vect{f}}_e$ is associated with the polarisation of
the bodies and $\hat{\vect{f}}_m$ is related to their magnetisation)
and
 \begin{equation}
\label{eq2.6}
\hat{H}_{AF}=-\sum_{m,n}\vect{d}_{mn}\sprod
 \hat{\vect{E}}(\vect{r}_{\!A})\hat{A}_{mn}
\end{equation}
($\vect{d}_{mn}$ $\!=$ $\!\langle m|\hat{\vect{d}}|n\rangle$,
electric-dipole transition matrix elements of the molecule;
$\vect{r}_{\!A}$, molecular centre-of-mass position; $\hat{A}_{mn}$
$\!=$ $\!|m\rangle\langle n|$, molecular flip operators) is the
molecule--field interaction Hamiltonian in electric-dipole
approximation. The electric field can be expressed in terms of the
bosonic variables according to
\begin{gather}
\label{eq2.7}
 \hat{\vect{E}}(\vect{r})
 =\int_0^{\infty}\!\dif\omega\,
 \underline{\hat{\vect{E}}}(\vect{r},\omega)+\operatorname{H.c.},\\
\label{eq2.7b}
 \underline{\hat{\vect{E}}}(\vect{r},\omega)=\sum_{\lambda={e},{m}}
 \int\dif^3r'\,\ten{G}_\lambda(\vect{r},\vect{r}',\omega)
 \!\cdot\!\hat{\vect{f}}_\lambda(\vect{r}',\omega),
\end{gather}
with the coefficients $\ten{G}_\lambda$ being related
to the classical Green tensor, $\ten{G}$, by
\begin{align}
\label{eq2.8}
&\ten{G}_e(\vect{r},\vect{r}',\omega)
 =\mi\,\frac{\omega^2}{c^2}
 \sqrt{\frac{\hbar}{\pi\varepsilon_0}\,
 \operatorname{Im}\varepsilon(\vect{r}',\omega)}\,
 \ten{G}(\vect{r},\vect{r}',\omega),\\[1ex]
\label{eq2.9}
&\ten{G}_m(\vect{r},\vect{r}',\omega)
 =\mi\,\frac{\omega}{c}
 \sqrt{\frac{\hbar}{\pi\varepsilon_0}\,
 \frac{\operatorname{Im}\mu(\vect{r}',\omega)}
 {|\mu(\vect{r}',\omega)|^2}}
 \bigl[\vect{\nabla}'
 \!\!\times\!\ten{G}(\vect{r}',\vect{r},\omega)
 \bigr]^{\trans}.
\end{align}
For a given environment of macroscopic bodies, described by their
linear, local and isotropic relative permittivity
$\varepsilon(\vect{r},\omega)$ and permeability
$\mu(\vect{r},\omega)$, the Green tensor is uniquely defined by the
differential equation
\begin{equation}
\label{eq2.10}
\left[\bm{\nabla}\times
\frac{1}{\mu(\vect{r},\omega)}\bm{\nabla}\times
 \,-\,\frac{\omega^2}{c^2}\,\varepsilon(\vect{r},\omega)\right]
 \ten{G}(\vect{r},\vect{r}',\omega)
 =\bm{\delta}(\vect{r}-\vect{r}')
\end{equation}
together with the boundary condition
\begin{equation}
\label{eq2.11}
\ten{G}(\vect{r},\vect{r}',\omega)\to \tenszero
\quad\mbox{for }|\vect{r}-\vect{r}'|\to\infty.
\end{equation}
The above definitions imply the useful integral relation
\cite{0002,0696}
\begin{multline}
\label{eq2.12}
\sum_{\lambda={e},{m}}\int\dif^3 s\,
 \ten{G}_\lambda(\vect{r},\vect{s},\omega)\!\cdot\!
 \ten{G}^{+}_\lambda\!(\vect{r}',\vect{s},\omega)\\
=\frac{\hbar\mu_0}{\pi}\,\omega^2\operatorname{Im}
 \ten{G}(\vect{r},\vect{r}',\omega).
\end{multline}

In thermal equilibrium at uniform temperature $T$, the electromagnetic
field may be described by the density matrix
\begin{equation}
\label{eq2.13}
\hat{\rho}_T
=\frac{\me^{-\hat{H}_\mathrm{F}/(k_\mathrm{B}T)}}
{\trace\bigl[\me^{-\hat{H}_\mathrm{F}/(k_\mathrm{B}T)}\bigr]}
\end{equation}
($k_\mathrm{B}$, Boltzmann constant). Thermal averages
$\langle\ldots\rangle$ $\!=$ $\!\trace[\ldots\hat{\rho}_T]$ of the
bosonic variables are thus given by
\begin{gather}
\label{eq2.14}
\bigl\langle\hat{\vect{f}}_\lambda(\vect{r},\omega)
 \bigr\rangle=\veczero
 =\bigl\langle\hat{\vect{f}}_\lambda^\dagger(\vect{r},\omega)
 \bigr\rangle,\\
\label{eq2.15}
\bigl\langle\hat{\vect{f}}_\lambda(\vect{r},\omega)
\hat{\vect{f}}_{\lambda'}(\vect{r}',\omega')\bigr\rangle=\tenszero
 =\bigl\langle\hat{\vect{f}}_\lambda^\dagger(\vect{r},\omega)
\hat{\vect{f}}_{\lambda'}^\dagger(\vect{r}',\omega')
\bigr\rangle,\\
\label{eq2.16}
\bigl\langle\hat{\vect{f}}_\lambda^\dagger(\vect{r},\omega)
\hat{\vect{f}}_{\lambda'}(\vect{r}',\omega')\bigr\rangle
=n(\omega)\delta_{\lambda\lambda'}
 \bm{\delta}(\vect{r}-\vect{r}')\delta(\omega-\omega'),\\
\bigl\langle\hat{\vect{f}}_\lambda(\vect{r},\omega)
\hat{\vect{f}}_{\lambda'}^\dagger(\vect{r}',\omega')\bigr\rangle
\hspace{14ex}\nonumber\\
\hspace{14ex}=[n(\omega)+1]\delta_{\lambda\lambda'}
 \bm{\delta}(\vect{r}-\vect{r}')\delta(\omega-\omega')
\label{eq2.17}
\end{gather}
where
\begin{equation}
\label{eq3.8}
n(\omega)=\frac{\sum_k k\me^{-k\hbar\omega/(k_\mathrm{B}T)}}
{\sum_k \me^{-k\hbar\omega/(k_\mathrm{B}T)}}
=\frac{1}{\me^{\hbar\omega/(k_\mathrm{B}T)}-1}
\end{equation}
is the average thermal photon number. Recalling
definitions~(\ref{eq2.7}) and (\ref{eq2.7b}), the statistical
properties of the electric field are found to be given by
\begin{gather}
\label{eq2.18}
\bigl\langle\underline{\hat{\vect{E}}}(\vect{r},\omega)
 \bigr\rangle=\veczero
 =\bigl\langle\underline{\hat{\vect{E}}}{}^\dagger(\vect{r},\omega)
 \bigr\rangle,\\
\label{eq2.19}
\bigl\langle\underline{\hat{\vect{E}}}(\vect{r},\omega)
\underline{\hat{\vect{E}}}(\vect{r}',\omega')\bigr\rangle=\tenszero
 =\bigl\langle\underline{\hat{\vect{E}}}{}^\dagger(\vect{r},\omega)
\underline{\hat{\vect{E}}}{}^\dagger(\vect{r}',\omega')
\bigr\rangle,\\
\label{eq2.20}
\bigl\langle\underline{\hat{\vect{E}}}{}^\dagger(\vect{r},\omega)
\underline{\hat{\vect{E}}}(\vect{r}',\omega')\bigr\rangle
\hspace{22ex}\nonumber\\
\hspace{10ex}=\frac{\hbar\mu_0}{\pi}n(\omega)\omega^2
 \operatorname{Im}\ten{G}(\vect{r},\vect{r}',\omega)
 \delta(\omega-\omega'),\\
\label{eq2.21}
\bigl\langle\underline{\hat{\vect{E}}}(\vect{r},\omega)
\underline{\hat{\vect{E}}}{}^\dagger(\vect{r}',\omega')\bigr\rangle
\hspace{32ex}\nonumber\\
\hspace{4ex}=\frac{\hbar\mu_0}{\pi}[n(\omega)+1]\omega^2
 \operatorname{Im}\ten{G}(\vect{r},\vect{r}',\omega)
 \delta(\omega-\omega')
\end{gather}
where we have made use of the integral relation~(\ref{eq2.12}). Note
that these relations are in accordance with the
fluctuation--dissipation theorem \cite{0751},
\begin{multline}
\label{eq2.22}
\bigl\langle{\textstyle\frac{1}{2}}
\bigl[\underline{\hat{\vect{E}}}(\vect{r},\omega)
\underline{\hat{\vect{E}}}{}^\dagger(\vect{r}',\omega')
+\underline{\hat{\vect{E}}}{}^\dagger(\vect{r}',\omega')
\underline{\hat{\vect{E}}}(\vect{r},\omega)\bigr]\bigr\rangle\\
=\frac{\hbar\mu_0}{\pi}
 \bigl[n(\omega)+{\textstyle\frac{1}{2}}\bigr]\omega^2
 \operatorname{Im}\ten{G}(\vect{r},\vect{r}',\omega)
 \delta(\omega-\omega'),
\end{multline}
where the thermal photon energy is given by
\begin{equation}
\label{eq2.23}
\hbar\omega \bigl[n(\omega)+{\textstyle\frac{1}{2}}\bigr]\to
\begin{cases}
\frac{1}{2}\hbar\omega
\quad\mbox{for }k_\mathrm{B}T\ll\hbar\omega,\\
k_\mathrm{B}T
\quad\mbox{for }k_\mathrm{B}T\gg\hbar\omega
\end{cases}
\end{equation}
in the zero- and high-temperature limits, respectively.


\section{Internal molecular dynamics}
\label{Sec3}

Consider a molecule which is prepared at initial time $t$ $\!=$ $\!0$
in an arbitrary internal state, represented by its internal density
matrix $\hat{\sigma}(0)$. The environment of the molecule is initially
taken to be at uniform temperature $T$, so that the electromagnetic
field is in a thermal state $\hat{\rho}(0)$ $\!=$ $\!\hat{\rho}_T$.

The internal molecular dynamics can be determined by solving the
coupled equations
\begin{multline}
\label{eq3.1}
\dot{\hat{A}}_{mn}=\frac{\mi}{\hbar}\bigl[\hat{H},\hat{A}_{mn}\bigr]
 =\mi\omega_{mn} \hat{A}_{mn}\\
 +\frac{\mi}{\hbar}\sum_k\int_0^\infty\dif\omega\biggl[
 \bigl(\vect{d}_{nk}\hat{A}_{mk}-\vect{d}_{km}\hat{A}_{kn}\bigr)\sprod
 \underline{\hat{\vect{E}}}(\vect{r}_{\!A},\omega)\\
+\underline{\hat{\vect{E}}}{}^\dagger(\vect{r}_{\!A},\omega)\sprod
 \big(\vect{d}_{nk}\hat{A}_{mk}-\vect{d}_{km}\hat{A}_{kn}\big)\biggr],
\end{multline}
and
\begin{multline}
\label{eq3.1b}
\dot{\hat{\vect{f}}}_\lambda(\vect{r},\omega)
=\frac{\mi}{\hbar}\bigl[\hat{H},
 \hat{\vect{f}}_\lambda(\vect{r},\omega)\bigr]\\
 =-\mi\omega\hat{\vect{f}}_\lambda(\vect{r},\omega)
 +\frac{\mi}{\hbar}\sum_{m,n}\vect{d}_{mn}\sprod
 \ten{G}_\lambda^\ast(\vect{r}_{\!A},\vect{r},\omega)\hat{A}_{mn},
\end{multline}
as implied by the Hamiltonian~(\ref{eq2.1}) together with
Eqs.~(\ref{eq2.2}), (\ref{eq2.3}) and (\ref{eq2.6}). The
electromagnetic field can be eliminated by formally solving
Eq.~(\ref{eq3.1b}) and substituting the result into Eq.~(\ref{eq3.1}).
For weak molecule--field coupling, the Markov approximation may then
be employed to show that the dynamics of the internal density matrix
of the molecule $\hat{\sigma}$ is given by the equations
(App.~\ref{AppA})
\begin{eqnarray}
\label{eq3.2}
\dot{\sigma}_{nn}(t)
&\!=&\!-\Gamma_n\sigma_{nn}(t)
 +\sum_k\Gamma_{kn}\sigma_{kk}(t),\\
\label{eq3.3}
\dot{\sigma}_{mn}(t)
&\!=&\!\bigl[-\mi\tilde{\omega}_{mn}
 -{\textstyle\frac{1}{2}}(\Gamma_m+\Gamma_n)\bigr]
 \sigma_{mn}(t)\nonumber\\
&\!&\!\mbox{for }m\neq n
\end{eqnarray}
($\sigma_{mn}$ $\!=$ $\!\langle m|\hat{\sigma}|n\rangle$ $\!=$
$\!\langle\hat{A}_{nm}\rangle$). Here, the total loss rate $\Gamma_n$
of a level $n$ is given by
\begin{equation}
\label{eq3.4}
\Gamma_n=\Gamma_n(\vect{r}_{\!A})
 =\sum_k \Gamma_{nk},
\end{equation}
and the individual intra-molecular transition rates $\Gamma_{nk}$ from
level $n$ to level $k$ read
\begin{eqnarray}
\label{eq3.5}
\Gamma_{nk}&\!=&\!\Gamma_{nk}(\vect{r}_{\!A})
 \equiv\Gamma_{nk}^0+\Gamma_{nk}^T\nonumber\\
&\!=&\!\frac{2\mu_0}{\hbar}\,
 \tilde{\omega}_{nk}^2
 \vect{d}_{nk}\sprod\operatorname{Im}
 \ten{G}(\vect{r}_{\!A},\vect{r}_{\!A},|\tilde{\omega}_{nk}|)
 \sprod\vect{d}_{kn}\nonumber\\
&\!&\!\times\{\Theta(\tilde{\omega}_{nk})[n(\tilde{\omega}_{nk})+1]
 +\Theta(\tilde{\omega}_{kn})n(\tilde{\omega}_{kn})\}
\end{eqnarray}
[$\Theta(z)$, unit step function] where
\begin{equation}
\label{eq3.6}
\Gamma_{nk}^0=\frac{2\mu_0}{\hbar}\,
 \tilde{\omega}_{nk}^2
\Theta(\tilde{\omega}_{nk})
 \vect{d}_{nk}\sprod\operatorname{Im}
 \ten{G}(\vect{r}_{\!A},\vect{r}_{\!A},\tilde{\omega}_{nk})
 \sprod\vect{d}_{kn}
\end{equation}
and
\begin{multline}
\label{eq3.7}
\Gamma_{nk}^T=\frac{2\mu_0}{\hbar}\,
 \tilde{\omega}_{nk}^2\vect{d}_{nk}\sprod\operatorname{Im}
 \ten{G}(\vect{r}_{\!A},\vect{r}_{\!A},|\tilde{\omega}_{nk}|)
 \sprod\vect{d}_{kn}\\
 \times[\Theta(\tilde{\omega}_{nk})n(\tilde{\omega}_{nk})
 +\Theta(\tilde{\omega}_{kn})n(\tilde{\omega}_{kn})]
\end{multline}
denote the zero-point and thermal contributions to these rates [recall
Eq.~(\ref{eq3.8})].

The intra-molecular transition rates depend on the shifted molecular
transition frequencies
\begin{equation}
\label{eq3.9}
\tilde{\omega}_{mn}=\tilde{\omega}_{mn}(\vect{r}_{\!A})
=\omega_{mn}+\delta\omega_m-\delta\omega_n,
\end{equation}
where the frequency shift
\begin{equation}
\label{eq3.10}
\delta\omega_n=\delta\omega_n(\vect{r}_{\!A})
 =\sum_k \delta\omega_{nk},
\end{equation}
of a given level $n$ has contributions
\begin{eqnarray}
\label{eq3.11}
\delta\omega_{nk}&\!=&\!\delta\omega_{nk}(\vect{r}_{\!A})
\equiv\delta\omega_{nk}^0+\delta\omega_{nk}^T\nonumber\\
&\!=&\!\frac{\mu_0}{\pi\hbar}\,
 \mathcal{P}\int_0^\infty\dif\omega\,\omega^2\biggl\{
 \vect{d}_{nk}\sprod\operatorname{Im}
 \ten{G}^{(1)}(\vect{r}_{\!A},\vect{r}_{\!A},\omega)\sprod
 \vect{d}_{kn}\nonumber\\
&\!&\!\times\biggl[\frac{n(\omega)+1}
 {\tilde{\omega}_{nk}-\omega}
 +\frac{n(\omega)}
 {\tilde{\omega}_{nk}+\omega}\biggr]\nonumber\\
&\!&\!+\,\frac{\omega|\vect{d}_{nk}|^2}{6\pi c}\,
 \biggl[\frac{n(\omega)}
 {\tilde{\omega}_{nk}-\omega}
 +\frac{n(\omega)}
 {\tilde{\omega}_{nk}+\omega}\biggr]\biggr\}
\end{eqnarray}
($\mathcal{P}$, principal value) due to all other levels $k$, which
can again be separated into their zero-point and thermal parts,
\begin{eqnarray}
\label{eq3.12}
\delta\omega_{nk}^0&\!=&\!\delta\omega_{nk}^0(\vect{r}_{\!A})
 \nonumber\\
&\!=&\!\frac{\mu_0}{\pi\hbar}\,
 \mathcal{P}\int_0^\infty\dif\omega\,\omega^2\,
 \frac{\vect{d}_{nk}\sprod\operatorname{Im}
 \ten{G}^{(1)}(\vect{r}_{\!A},\vect{r}_{\!A},\omega)\sprod
 \vect{d}_{kn}}{\tilde{\omega}_{nk}-\omega}\nonumber\\
\end{eqnarray}
and
\begin{eqnarray}
\label{eq3.13}
\delta\omega_{nk}^T&\!=&\!\delta\omega_{nk}^T(\vect{r}_{\!A})
 \nonumber\\
&\!=&\!\frac{\mu_0}{\pi\hbar}\,
 \mathcal{P}\int_0^\infty\dif\omega\,\omega^2
 \vect{d}_{nk}\sprod\operatorname{Im}
 \ten{G}(\vect{r}_{\!A},\vect{r}_{\!A},\omega)\sprod
 \vect{d}_{kn}\nonumber\\
&\!&\!\times\biggl[\frac{n(\omega)}{\tilde{\omega}_{nk}-\omega}
 +\frac{n(\omega)}{\tilde{\omega}_{nk}+\omega}\biggr]\,,
\end{eqnarray}
respectively. Here, $\ten{G}^{(1)}$ denotes the scattering part of
the Green tensor according to the decomposition
\begin{equation}
\label{eq3.13b}
\ten{G}(\vect{r},\vect{r}',\omega)
=\ten{G}^{(0)}(\vect{r},\vect{r}',\omega)
+\ten{G}^{(1)}(\vect{r},\vect{r}',\omega)
\end{equation}
where the imaginary part of the bulk (free-space) part is given by
\cite{0003}
\begin{equation}
\label{eq3.22}
\operatorname{Im}\ten{G}^{(0)}(\vect{r},\vect{r},\omega)
=\frac{\omega}{6\pi c}\,\ten{I}
\end{equation}
($\ten{I}$, unit tensor). The free-space zero-point frequency shifts
associated with $\ten{G}^{(0)}$, i.e., the free-space Lamb shifts, are
included in the bare transition frequencies $\omega_{mn}$ since they
are determined experimentally in free space. The Green tensor being
analytic in the upper half of the complex frequency plane, one can
employ contour-integral techniques to rewrite the frequency-shift
contributions as
\begin{multline}
\label{eq3.13c}
\delta\omega_{nk}
 =-\frac{\mu_0}{\hbar}\,
 \tilde{\omega}_{nk}^2\vect{d}_{nk}\sprod\operatorname{Re}
 \ten{G}^{(1)}(\vect{r}_{\!A},\vect{r}_{\!A},\tilde{\omega}_{nk})
 \sprod\vect{d}_{kn}\\
\times\{\Theta(\tilde{\omega}_{nk})[n(\tilde{\omega}_{nk})+1]
 -\Theta(\tilde{\omega}_{kn})n(\tilde{\omega}_{kn})\}\\
+\frac{2\mu_0k_\mathrm{B}T}{\hbar^2}\sum_{N=0}^\infty
 (1-\delta_{N0})\xi_N^2\tilde{\omega}_{kn}\\
\times\frac{\vect{d}_{nk}\sprod
 \ten{G}^{(1)}(\vect{r}_{\!A},\vect{r}_{\!A},\xi_N)
 \sprod\vect{d}_{kn}}
 {\tilde{\omega}_{kn}^2+\xi_N^2}\\
+\frac{\mu_0|\vect{d}_{nk}|^2}{6\pi^2c\hbar}
 \mathcal{P}\int_0^\infty\dif\omega\,\omega^3
 \biggl[\frac{n(\omega)}
 {\tilde{\omega}_{nk}-\omega}
 +\frac{n(\omega)}
 {\tilde{\omega}_{nk}+\omega}\biggr]
\end{multline}
[note that $\operatorname{Re}\ten{G}(\vect{r},\vect{r}',-\omega)$
$\!=$ $\operatorname{Re}\ten{G}(\vect{r},\vect{r}',\omega)$ for
real $\omega$] with Matsubara frequencies
\begin{equation}
\label{eq3.13d}
\xi_N=\frac{2\pi k_\mathrm{B}T}{\hbar}\,N,\qquad N=0,1,\ldots
\end{equation}
When neglecting the frequency shifts, the transition
rates~(\ref{eq3.4})--(\ref{eq3.7}) obviously reduce to the well-known
results given, e.g., in Ref.~\cite{0041}.

It is worth noting that the internal molecular dynamics described by
Eqs.~(\ref{eq3.2}) and (\ref{eq3.3}) obeys probability conservation,
\begin{multline}
\label{eq3.14}
\frac{\dif}{\dif t}\,\trace\,\hat{\sigma}(t)
 =\sum_n\dot{\sigma}_{nn}(t)\\
 =-\sum_{n,k} \Gamma_{nk}\sigma_{nn}(t)
 +\sum_{n,k}\Gamma_{kn}\sigma_{kk}(t)
 =0,
\end{multline}
where we have used Eq.~(\ref{eq3.4}). From the property
\begin{equation}
\label{eq3.15}
\Gamma_{nk}=\me^{\hbar\tilde{\omega}_{nk}/(k_\mathrm{B}T)}
 \Gamma_{kn}
\end{equation}
of the transition rates [see Eq.~(\ref{eq3.5})], it follows that in
the long-time limit the molecule reaches a thermal state as its steady
state
\begin{equation}
\label{eq3.16}
\hat{\sigma}(t\to\infty)=\hat{\sigma}
_T
=\frac{\me^{-\sum_n\tilde{E}_n|n\rangle\langle n|/(k_\mathrm{B}T)}}
 {\trace\bigl[
 \me^{-\sum_n\tilde{E}_n|n\rangle\langle n|/(k_\mathrm{B}T)}\bigr]}
\end{equation}
with
\begin{equation}
\label{eq3.17}
\tilde{E}_n=\tilde{E}_n(\vect{r}_{\!A})=E_n+\hbar\delta\omega_n
\end{equation}
denoting the shifted molecular eigenenergies. This can be verified by
noting that for this state the internal molecular evolution as
given by Eqs.~(\ref{eq3.2}) and (\ref{eq3.3}) becomes static,
\begin{align}
\label{eq3.18}
&\dot{\sigma}_{nn}(t\to\infty)=-\Gamma_n\sigma
 _{nn,T}
 +\sum_k\Gamma_{kn}\sigma
 _{kk,T}\nonumber\\
&\quad=-\sum_k \Gamma_{nk}\sigma
 _{nn,T}\nonumber\\
&\qquad+\sum_k\me^{-\hbar\tilde{\omega}_{nk}/(k_\mathrm{B}T)}
 \Gamma_n^k
 \me^{\hbar\tilde{\omega}_{nk}/(k_\mathrm{B}T)}\sigma
 _{nn,T}
 =0,\\
\label{eq3.19}
&\sigma_{mn}(t\to\infty)
 =\me^{\{-\mi\tilde{\omega}_{mn}-[\Gamma_m+\Gamma_n]/2\}(t-t_0)}
 \sigma
 _{mn,T}
 =0\nonumber\\
&\qquad\mbox{for }m\neq n.
\end{align}

According to Eqs.~(\ref{eq3.4}) and (\ref{eq3.5}), the heating
rate of a molecule prepared in its ground state $|0\rangle$ is given
(initially) by
\begin{multline}
\label{eq3.20}
\Gamma_0=\sum_k\Gamma_{0k}=\sum_k\Gamma_{0k}^T\\
=\frac{2\mu_0}{\hbar}\sum_k
 \tilde{\omega}_{k0}^2
 n(\tilde{\omega}_{k0})
\vect{d}_{0k}\sprod\operatorname{Im}
 \ten{G}(\vect{r}_{\!A},\vect{r}_{\!A},\tilde{\omega}_{k0})
 \sprod\vect{d}_{k0},
\end{multline}
due entirely to the absorption of thermal photons.


\section{Applications}
\label{Sec4}

The energy associated with electronic excitation of molecules is
typically large in comparison with thermal energy at room
temperature, i.e. \mbox{$\exp[-\hbar\omega_{n0}/(k_\mathrm{B}T)]$
$\!\ll$ $\!1$}, so according to Eq.~(\ref{eq3.15}), the fully
thermalised state effectively coincides with the electronic ground
state. This argument does not apply to the rotational and vibrational
excitations of polar molecules, which occur at much lower frequencies.
In this section, we study the ground-state heating rates $\Gamma_{0k}$
which provide a measure of the timescale on which this thermal
excitation of the rotational and vibrational states takes place. We
will assume that the frequency shifts induced by the environment are
small enough to justify putting $\tilde{\omega}_{mn}$ $\!=$
$\!\omega_{mn}$. In this case the thermal excitation rate from the
ground state to state $k$ becomes
\begin{multline}
\label{eq3.21b}
\Gamma_{0k}=\frac{2\mu_0}{\hbar}\,\omega_{k0}^2 n(\omega_{k0})
\vect{d}_{0k}\sprod\operatorname{Im}
 \ten{G}(\vect{r}_{\!A},\vect{r}_{\!A},\omega_{k0})
 \sprod\vect{d}_{k0}.
\end{multline}
This has the great virtue that the temperature appears only in the
thermal photon number $n(\omega_{k0})$ [recall Eq.~(\ref{eq3.8})],
while the position enters only through the Green tensor $\ten{G}$.
Therefore the dependence on temperature can be derived entirely from
considering the free-space case, while the position-dependence can be
understood completely from the behaviour at zero temperature.


\subsection{Molecules in free space}
\label{Sec4.1}
In free space, the Green tensor is given by Eq.~(\ref{eq3.22}), so the
molecular transition rates become
\begin{multline}
\label{eq3.23}
\Gamma_{nk}\equiv\Gamma_{nk}^{(0)}\equiv\Gamma_{nk}^0+\Gamma_{nk}^T\\
=\frac{|\omega_{nk}|^3|\vect{d}_{nk}|^2}{3\pi\hbar\varepsilon_0c^3}
 \{\Theta(\omega_{nk})[n(\omega_{nk})+1]
 +\Theta(\omega_{kn})n(\omega_{kn})\}
\end{multline}
with
\begin{equation}
\label{eq3.24}
\Gamma_{nk}^0
=\frac{\omega_{nk}^3|\vect{d}_{nk}|^2}{3\pi\hbar\varepsilon_0c^3}\,
 \Theta(\omega_{nk})
\end{equation}
and
\begin{equation}
\label{eq3.25}
\Gamma_{nk}^T
=\frac{|\omega_{nk}|^3|\vect{d}_{nk}|^2}{3\pi\hbar\varepsilon_0c^3}\,
 [\Theta(\omega_{nk})n(\omega_{nk})
 +\Theta(\omega_{kn})n(\omega_{kn})]\,.
\end{equation}
The total heating rate of a molecule initially prepared in its ground
state thus reads
\begin{equation}
\label{eq3.26}
\Gamma_0=\sum_k\Gamma_{0k}
=\sum_k\frac{\omega_{0k}^3|\vect{d}_{0k}|^2}
 {3\pi\hbar\varepsilon_0c^3}\,
 n(\omega_{k0}),
\end{equation}
in agreement with Ref.~\cite{0772}.

The ground-state heating rate of polar molecules will be dominated by
transitions to the adjacent excited rotational and vibrational states,
so we restrict our attention to these in the following. We calculate
the heating rates for the set of ground state polar molecules listed
in Tab.~\ref{Tab1}, which also gives the required molecular constants.
\begin{table*}
\begin{tabular}{llrlrlrlrlc}
\hline
Species\;&Ground state\;
&\multicolumn{2}{c}{$B_\mathrm{e}(\mathrm{GHz}$)}
&\multicolumn{2}{c}{$\omega_\mathrm{e}(\mathrm{THz})$}
&\multicolumn{2}{c}{$\mu_\mathrm{e}(10^{-30}\mathrm{Cm})$}
&\multicolumn{2}{c}{$\mu'_\mathrm{e}(10^{-21}\mathrm{C})$}
&$\;m(10^{-27}\mathrm{kg})^\ast\;$\\
\hline
\;LiH&\;\;$X^1\Sigma^+$&$222$&\cite{0817}
&$42.1$&\cite{0772}
&$19.6$&\cite{0818}
&$60.5$&\cite{0772}
&$1.46$\\
\;NH&\;\;$X^3\Sigma^-$&$500$&\cite{0798}
&$98.4$&\cite{0798}
&$5.15$&\cite{0799}
&${}^{\ast\ast}$&&$1.56$\\
\;OH${}^{\ast\ast\ast}$&\;\;$X^2\Pi$&$555$&\cite{0804}
&$112$&\cite{0807}
&$5.56$&\cite{0808}
&$17.9$&\cite{0809}&$1.57$\\
\;OD${}^{\ast\ast\ast}$&\;\;$X^2\Pi$&$300$&\cite{0805}
&$81.6$&\cite{0805}
&$5.51$&\cite{0808}
&${}^{\ast\ast}$&&$2.97$\\
\;CaF&\;\;$X^2\Sigma^+$&$10.5$&\cite{0819}
&$18.4$&\cite{0798}
&$10.2$&\cite{0820}
&$172$&\cite{0798}&$21.4$\\
\;BaF&\;\;$X^2\Sigma^+$&$6.30$&\cite{0772}
&$14.1$&\cite{0772}
&$11.7$&\cite{0772}
&$285$&\cite{0772}&$27.7$\\
\;YbF&\;\;$X^2\Sigma^+$&$7.20$&\cite{0821}
&$15.2$&\cite{0791}
&$13.1$&\cite{0821}
&$195$&\cite{0790}&$28.4$\\
\;LiRb&\;\;$X^1\Sigma^+$&$6.60$&\cite{0794}
&$5.55$&\cite{0772}
&$13.5$&\cite{0793}
&$21.4$&\cite{endnote,0772}&$10.8$\\
\;NaRb&\;\;$X^1\Sigma^+$&$2.03$&\cite{0794}
&$3.21$&\cite{0772}
&$11.7$&\cite{0793}
&$12.6$&\cite{0772}&$30.0$\\
\;KRb&\;\;$X^1\Sigma^+$&$1.15$&\cite{0796}
&$2.26$&\cite{0772}
&$0.667$&\cite{0793}
&$1.89$&\cite{0772}&$44.3$\\
\;LiCs&\;\;$X^1\Sigma^+$&$5.80$&\cite{0795}
&$4.92$&\cite{0772}
&$21.0$&\cite{0793}
&$28.4$&\cite{0772}&$11.1$\\
\;NaCs&\;\;$X^1\Sigma^+$&$17.7$&\cite{0795}
&$2.94$&\cite{0772}
&$19.5$&\cite{0793}
&$21.4$&\cite{0772}&$32.5$\\
\;KCs&\;\;$X^1\Sigma^+$&$92.8$&\cite{0795}
&$1.98$&\cite{0772}
&$8.61$&\cite{0793}
&$6.93$&\cite{0772}&$50.0$\\
\;RbCs&\;\;$X^1\Sigma^+$&$0.498$&\cite{0797}
&$1.48$&\cite{0772}
&$7.97$&\cite{0793}
&$4.41$&\cite{0772}&$86.0$\\
\end{tabular}
\caption{
\label{Tab1}
Properties of various diatomic radicals: electronic ground state,
rotation and vibration constants, dipole moment and its derivative at
equilibrium bond length, and reduced mass. For comparison with the
constants used in \cite{0772}, see \cite{endnote}.\\
${}^\ast$ Reduced masses are given on the basis of the atomic masses
(most abundant isotopes) of the molecular constituents as stated in
Ref.~\cite{0812}.\\
${}^{\ast\ast}$ For NH and OD, the electric-dipole matrix elements
for the transition between ground and first excited vibrational states
can be given as $|\vect{d}_{0k}|$ $\!=$
$\!1.80\!\times\!10^{-31}\mathrm{Cm}$ \cite{0799} and
$|\vect{d}_{0k}|$ $\!=$ $\!7.54\!\times\!10^{-32}\mathrm{Cm}$
\cite{0811}, respectively.\\
${}^{\ast\ast\ast}$ The spin-orbit coupling constants required for OH
and OD are $A$ $\!=$ $\!-4.189\,\mathrm{THz}$ \cite{0804} and $A$
$\!=$ $\!-4.174\,\mathrm{THz}$ \cite{0806}, respectively.
}
\end{table*}

We begin by considering rotational heating. To evaluate
Eq.~(\ref{eq3.26}) we will calculate the matrix elements of the
electric dipole operator using Hund's case (a) basis states
\cite{0813}. In this coupling scheme, the orbital angular momentum,
$\hat{\vect{L}}$, is strongly coupled to the internuclear axis, and so
is the electron spin, $\hat{\vect{S}}$, due to a strong spin-orbit
coupling. The total angular momentum is
$\hat{\vect{J}}=\hat{\vect{L}}+\hat{\vect{S}}+\hat{\vect{R}}$,
where $\hat{\vect{R}}$ is the angular momentum of the rotating nuclei
and is necessarily perpendicular to the internuclear axis. The
projections of $\hat{\vect{L}}$, $\hat{\vect{S}}$
and $\hat{\vect{J}}$ onto the internuclear axis are labelled by the
quantum numbers $\Lambda$, $\Sigma$ and $\Omega = \Lambda + \Sigma$.
The projection of $\hat{\vect{J}}$ onto the space-fixed $z$-axis is
$M$. The basis states are labelled by the quantum numbers $S$,
$\Lambda$, $\Sigma$, $\Omega$, $J$ and $M$.

For transitions between the rotational states, the matrix elements of
the electric dipole operator are
\begin{align}
\label{Eq:dipoleMe1}
&\vect{d}_{mn}=\langle\Omega JM|\hat{\vect{d}}|\Omega'J'M'\rangle
 =\mu_\mathrm{e}\langle \Omega JM|\hat{\vect{u}}|\Omega'J'M'\rangle
 \nonumber\\
&=\mu_\mathrm{e}\biggl[(u_{mn}^{-1}-u_{mn}^{+1})\frac{\vect{e}_{x}}{
\sqrt { 2 } }
 +(u_{mn}^{-1}+u_{mn}^{+1})\frac{\mi\,\vect{e}_{y}}{\sqrt{2}}
 +u_{mn}^{\,0}\vect{e}_{z}\biggr],
\end{align}
where $\mu_\mathrm{e}$ is the molecular dipole moment at the
equilibrium internuclear separation,
$\hat{\vect{u}}=\hat{\vect{r}}/|\hat{\vect{r}}|$, and
\begin{multline}
\label{Eq:dipoleMe2}
u_{mn}^{q}=(-1)^{M-\Omega}\sqrt{(2J+1)(2J'+1)}\\
 \times\begin{pmatrix}
  J & 1 & J' \\ -M & q & M'
 \end{pmatrix}
 \begin{pmatrix}
 J & 1 & J' \\ -\Omega  & 0 & \Omega'
 \end{pmatrix}.
\end{multline}
With this result, we obtain the selection rules for transitions
between the basis states: \mbox{$\Delta\Omega$ $\!=$ $\!0$},
\mbox{$\Delta J$ $\!=$ $\!0,\pm 1$}, and \mbox{$\Delta M$ $\!=$
$\!0,\pm 1$}. In this paper, we will not consider mixing of the
electronic ground state with other electronic states, which leads to
$\Lambda$-doubling, because the energy splitting that is induced is
very small compared with the rotational energies and so does not alter
any of our results. In this approximation, the states $|\pm\Omega J
M\rangle$ are degenerate, and since $\Delta \Omega = 0$ we can confine
our attention to the positive values of $\Omega$ only. While our
equations make it clear how to handle initial states of given $M'$, we
will consider the initial molecular state to be unpolarised, averaging
over the possible values of $M'$.

The majority of the molecules listed in Tab.~\ref{Tab1} have
\mbox{$\Lambda$ $\!=$ $\!0$} ground states. These molecules are best
described using Hund's coupling case (b) \cite{0813}. The spin is not
coupled to the internuclear axis and neither $\Sigma$ nor $\Omega$ is
defined. The rotational eigenenergies are
\begin{equation}
\label{eq3.26b}
E_N=h B_\mathrm{e} N(N+1),\quad N=0,1,\ldots
\end{equation}
where $B_\mathrm{e}$ is the rotational constant and $N$ is the
rotational quantum number,
$\hat{\vect{N}}=\hat{\vect{J}}-\hat{\vect{S}}$. The expansion of the
$\Sigma$ eigenstates in the case (a) basis is \cite{0813}
\begin{multline}
\label{expansion}
|S,N,J,M\rangle =\sum_{\Omega =-S}^S(-1)^{J-S}\sqrt{2N+1}\,\\
\times\begin{pmatrix}J&S&N\\ \Omega&-\Omega&0\end{pmatrix}
|\Omega,J,M\rangle
\end{multline}
Using Eqs.~(\ref{Eq:dipoleMe1}), (\ref{Eq:dipoleMe2}) and
(\ref{expansion}), summing over the possible final states and
averaging over initial states of different $M'$, we find
$\sum_{k}|\vect{d}_{0k}|^{2}$ $\!=$ $\!\mu_\mathrm{e}^{2}$ for
$^{1}\Sigma$, $^{2}\Sigma$, and $^{3}\Sigma$ molecules. For
$^{2}\Sigma$ molecules, the ground state $|N\!=\!0,J\!=\!1/2 \rangle$
can be excited either to $|N\!=\!1,J\!=\!1/2 \rangle$ or to
$|N\!=\!1,J\!=\!3/2 \rangle$, with branching ratios $1/3$ and $2/3$
respectively. The spin-rotation interaction lifts the degeneracy
between these states, but this splitting is very small and we do not
need to include it. For $^{3}\Sigma$ molecules, the ground state
$|N\!=\!0,J\!=\!1\rangle$ can be excited to the three states with
$N=1$ and $J=0,1,2$, with branching ratios $1/9$, $1/3$ and $5/9$
respectively. Again, we can neglect the small spin-rotation
interaction that lifts the degeneracy between the three states.

The electronic ground states of OH and OD are $^{2}\Pi$ states and,
for low values of $J$, are best described using Hund's coupling case
(a). The Hamiltonian describing the fine structure contains a
rotational part and a spin-orbit coupling, $\hat{H}_\mathrm{fs}$ $\!=$
$hA\hat{\vect{L}}\sprod\hat{\vect{S}}
+hB_\mathrm{e}(\hat{\vect{J}}-\hat{\vect{L}}-\hat{\vect{S}})^2$.
The rotational term couples states of the same $J$ but different
$|\Omega|$. Writing the matrix elements of the Hamiltonian as
$m_{\Omega,\Omega'}=\langle\Omega JM
|\hat{H}_\mathrm{fs}|\Omega'JM\rangle$ we
have \cite{0813}
\begin{align}
m_{\substack{3/2,3/2\\ 1/2,1/2}}&=\pm hA/2 +
hB_\mathrm{e}[J(J+1)-3/4 \mp 1],\\
m_{3/2,1/2} &= -hB_\mathrm{e}\sqrt{(J+3/2)(J-1/2)}.
\end{align}
Diagonalizing this Hamiltonian gives a pair of energy eigenvalues for
each value of $J>1/2$,
\begin{equation}
\label{eq3.26d5}
E_{J}=hB_\mathrm{e}[(J+1/2)^{2}-1 \pm\mathcal{Q}/2],
\end{equation}
where
\begin{equation}
\mathcal{Q}=\sqrt{4(J+1/2)^{2}+A/B_\mathrm{e}(A/B_\mathrm{e}-4)}
\end{equation}

We will use the labels $F_1$ and $F_2$ to denote the states of lower
and higher energy, respectively. For the low-$J$ levels of OH and OD,
the mixing of $\Omega$ states is small because $|A|$ is considerably
larger than $B_\mathrm{e} J$. Recalling that $A$ is negative for these
molecules, we can then identify $F_{1}$ as having predominantly
$^2\Pi_{3/2}$ character, and $F_{2}$ as predominantly $^2\Pi_{1/2}$.
For $J=1/2$ there is only one level, which is of pure $\Omega=1/2$
character. The eigenstates are
\begin{align}
\label{eq3.26d7}
|F_{1},J,M\rangle=&c_+(J)|1/2, J,M\rangle
 +c_-(J)|3/2,J,M\rangle,\nonumber\\
&\qquad J=3/2,5/2,\ldots,\\
\label{eq3.26d6}
|F_{2}J,M\rangle=&c_+(J)|3/2,J,M\rangle
 -c_-(J)|1/2,J,M\rangle, \nonumber\\
&\qquad J=1/2,3/2\ldots
\end{align}
where
\begin{equation}
\label{eq3.26d8}
c_\pm(J)=\sqrt{1/2\pm(A/B_\mathrm{e}-2)/(2{\cal Q})}\,.
\end{equation}
Using the selection rules between the basis states, we see that the
possible transitions out of the molecular ground state
$|F_{1},J\!=\!3/2\rangle$, are those to the states
\mbox{(a)$|F_{1},J\!=\!5/2\rangle$},
\mbox{(b)$|F_{2},J\!=\!1/2\rangle$},
\mbox{(c) $|F_{2},J\!=\!3/2\rangle$}
and \mbox{(d) $|F_{2},J\!=\!5/2\rangle$}.
Applying Eqs.~(\ref{Eq:dipoleMe1}) and (\ref{Eq:dipoleMe2}) to each of
these four transitions, summing over the $M$ sublevels in the final
state, and averaging over the $M'$ sublevels in the initial state, we
obtain
\begin{align}
&\sum_{k(a)}|\vect{d}_{0k}|^{2}
 =\Bigl[{\textstyle\frac{3}{5}}c_+^2(3/2)c_+^2(5/2)
 +{\textstyle\frac{2}{5}}c_-^2(3/2)c_-^2(5/2)\nonumber\\
&\qquad+{\textstyle\frac{6}{5}}{\textstyle\sqrt{\frac{2}{3}}}
 c_+(3/2)c_+(5/2)c_-(3/2)c_-(5/2)\Bigr]
 \mu_\mathrm{e}^2,\\
&\sum_{k(b)}|\vect{d}_{0k}|^{2}
 ={\textstyle\frac{1}{3}}c_+^2(3/2)\mu_\mathrm{e}^2,\\
&\sum_{k(c)}|\vect{d}_{0k}|^{2}
 ={\textstyle\frac{4}{15}}c_+^2(3/2)c_-^2(3/2)
 \mu_\mathrm{e}^2,
\end{align}
\begin{align}
&\sum_{k(d)}|\vect{d}_{0k}|^{2}
 =\Bigl[{\textstyle\frac{3}{5}}c_+^2(3/2)c_-^2(5/2)
 +{\textstyle\frac{2}{5}}c_-^2(3/2)c_+^2(5/2)\nonumber\\
&\qquad-{\textstyle\frac{6}{5}}{\textstyle\sqrt{\frac{2}{3}}}
 c_+(3/2)c_-(3/2)c_+(5/2)c_-(5/2)\Bigr]
 \mu_\mathrm{e}^2.
\end{align}

\begin{figure}[!t!]
\includegraphics[width=\linewidth]{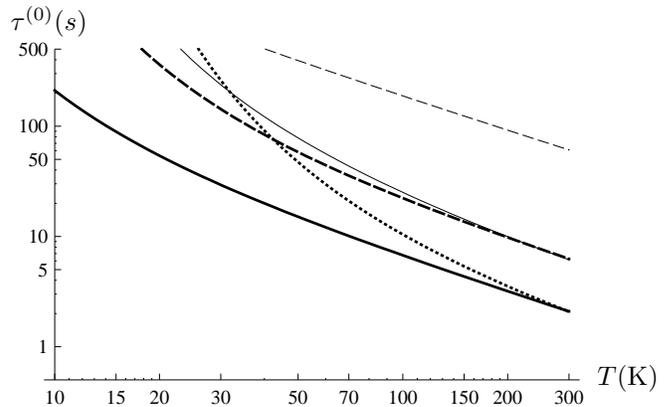}
\caption{
\label{Fig1}
Free-space lifetimes of the ground state against rotational heating as
a function of environment temperature for LiH (thick solid line), NH
(thick dashed line), OH(a) (thick dotted line), OD(a) (thin solid
line) and KCs (thin dashed line).}
\end{figure}

With these preparations, we can now evaluate the rates for free-space
rotational heating out of the ground-state, for the molecules listed
in Tab.~\ref{Tab1}. The lifetimes, $\tau^{(0)}$
$\!=$ $\!(\Gamma^{(0)})^{-1}$, are given in Tab.~\ref{Tab2} for
environmental temperatures of $293\,\mathrm{K}$ and $77\,\mathrm{K}$.
Since there is little variation of the dipole moment, the lifetime is
mainly determined by the power of the thermal spectrum at the
transition frequency. Apart from the weakest transitions in OH and OD
all these lines lie on the low side of the peak frequency in the
thermal spectrum, which is $17\,\mathrm{THz}$ at $293\,\mathrm{K}$ or
$5\,\mathrm{THz}$ at $77\,\mathrm{K}$. Note that the rotational
constant is roughly given by \mbox{$B_\mathrm{e}$ $\!\approx$
$\!\hbar/(4\pi mR_\mathrm{e}^2)$} where $R_\mathrm{e}$ is the
equilibrium internuclear separation and $m=m_1m_2/(m_1+m_2)$ is the
reduced mass, so as a rule of thumb, rotational heating is most severe
for the light molecules.  Strong heating is seen for LiH, NH, OH, OD,
whose lifetimes are in the range of 2--6 seconds. For KCs and NaCs the
heating is much less severe, and for the rest it is negligible for
most practical purposes. Tab.~\ref{Tab2} also shows that rotational
heating of OH and OD is dominated by transition (a), with the other
transitions providing small corrections to the heating rate, even
though they are at higher frequencies. This behaviour is due to the
exceedingly small transition dipole moments of the latter transitions.
The rotational excitation lifetimes of all these molecules can be
extended by going to lower environmental temperatures.
Figure~\ref{Fig1} illustrates this temperature-dependence in the light
molecules LiH, NH, OH, OD and KCs.
\begin{table}
\begin{tabular}{ccccc}
\hline
&&&\multicolumn{2}{c}{$\tau^{(0)}(\mathrm{s})$}\\
Species&$\frac{\omega_{0k}}{2\pi}(\mathrm{GHz})$
&$\frac{\sum_{k}|\vect{d}_{0k}|^{2}}{\mu_\mathrm{e}^{2}}$
&$293\,\mathrm{K}$&$77\,\mathrm{K}$\\
\hline
LiH&$\;444\;$&\;1\;&$\;2.1\;$
&$\;9.1\;$\\
NH&$\;999\;$&\;1\;&$\;6.4\;$
&$\;31\;$\\
OH&$\;\;$&$\;\;$
&$\;2.1\;$&$\;17\;$\\
(a)&$\;2.51\!\times\!10^{3}$\;&\;0.405\;&$\;2.4\;$
&$\;18\;$\\
(b)&$\;3.80\!\times\!10^{3}\;$&\;0.00999\;&$\;49\;$
&$\;550\;$\\
(c)&$\;5.64\!\times\!10^{3}\;$&\;0.00775\;&$\;34\;$
&$\;720\;$\\
(d)&$\;8.67\!\times\!10^{3}\;$&\;0.00124\;&$\;120\;$
&$\;8,400\;$\\
OD&$\;\;$&\;\;&$\;6.3\;$
&$\;37\;$\\
(a)&$\;1.41\!\times\!10^{3}\;$&\;0.402\;&$\;7.2\;$
&$\;39\;$\\
(b)&$\;3.93\!\times\!10^{3}\;$&\;0.00381\;&$\;120\;$
&$\;1,\!400\;$\\
(c)&$\;4.89\!\times\!10^{3}\;$&\;0.00302\;&$\;110\;$
&$\;1,\!800\;$\\
(d)&$\;6.48\!\times\!10^{3}\;$&\;0.000636\;&$\;340\;$
&$\;10,\!000\;$\\
CaF&$\;21.0\;$&\;1\;&$3,\!400$
&$13,\!000$\\
BaF&$\;12.6\;$&\;1\;&$\;7,\!200\;$
&$\;28,\!000\;$\\
YbF&$\;14.4\;$&\;1\;&$\;4,\!400\;$
&$\;17,\!000\;$\\
LiRb&$\;13.2\;$&\;1\;&$\;4,\!900\;$
&$\;19,\!000\;$\\
NaRb&$\;4.05\;$&\;1\;&$\;70,\!000\;$
&$\;260,\!000\;$\\
KRb&$\;2.30\;$&\;1\;&$\;6.7\!\times\!10^{7}\;$
&$\;2.5\!\times\!10^{8}\;$\\
LiCs&$\;11.6\;$&\;1\;&$\;2,\!600\;$
&$\;10,\!000\;$\\
NaCs&$\;35.5\;$&\;1\;&$\;330\;$
&$\;13,\!000\;$\\
KCs&$\;186\;$&\;1\;&$\;62\;$
&$\;250\;$\\
RbCs&$\;0.995\;$&\;1\;&$\;2.5\!\times\!10^{6}\;$
&$\;9.5\!\times\!10^{6}\;$\\
\end{tabular}
\caption{
\label{Tab2}
Free-space lifetimes for rotational heating out of the ground state at
$293\,\mathrm{K}$ and $77\,\mathrm{K}$. For OH and OD, the effects of
the transitions (a)--(d) (see main text) are also shown separately.
Also given are the frequency and the square of the dipole matrix
element for each transition. For comparison with the results of
\cite{0772}, see \cite{endnote}.
}
\end{table}

Let us next turn our attention to vibrational heating. To a good
approximation, the  vibrational eigenenergies of the deeply-bound
states of a diatomic molecule are
\begin{equation}
\label{eq3.26f}
E_v=h\omega_\mathrm{e}
 \bigl(v+{\textstyle\frac{1}{2}}\bigr),\quad v=0,1,\ldots
\end{equation}
where $\omega_\mathrm{e}$ is the vibrational constant and $v$ is the
vibrational quantum number. The corresponding eigenstates are
\begin{equation}
\label{eq3.26g}
\langle q|v\rangle=\left(\frac{\alpha}{\pi}\right)^{1/4}\,
 \frac{1}{\sqrt{2^vv!}}\,H_v\bigl(\sqrt{\alpha}\,q\bigr)
 \me^{-\frac{1}{2}\alpha q^2},
\end{equation}
where $q$ $\!=$ $R\!-\!R_\mathrm{e}$, $R$ being the internuclear
separation, $H_n$ are the Hermite polynomials and $\alpha=2\pi
m\omega_\mathrm{e}/\hbar$. Expanding the electric-dipole operator in a
Taylor series about the equilibrium separation, $\hat{\vect{d}}$
$\!\approx$ $\!(\mu_\mathrm{e}+\mu'_\mathrm{e}\hat{q})\hat{\vect{u}}$,
and recognizing that the first term cannot couple different
vibrational states, we write the matrix elements for rovibrational
transitions in the form
\begin{equation}
\langle v\,\Omega JM|\hat{\vect{d}}|v'\,\Omega'J M'\rangle
 =\mu_\mathrm{e}'\langle\Omega JM|\hat{\vect{u}}|\Omega'J'M'\rangle
 \langle v|\hat{q}|v'\rangle.
\end{equation}
We see from this equation that the rovibrational transitions must
satisfy the same rotational selection rules as already given above,
and that to leading order in $q$, the vibrational selection rule is
$\Delta v$ $\!=$ $\!\pm 1$. For transitions between $v'=0$ and $v=1$
we have
\begin{equation}
\langle v=1|\hat{q}|v'=0\rangle=\frac{1}{\sqrt{2\alpha}}
 =\sqrt{\frac{\hbar}{4\pi m\omega_\mathrm{e}}}\,.
\end{equation}
We neglect the contribution of rotational energy to the transition
frequency since it is typically smaller than the vibrational energy by
two orders of magnitude. This means that we can simply add up the
contributions of transitions (a)--(d) in calculating the transition
dipole moments for OH and OD. Thus we obtain,
\begin{equation}
\sum_{k}|\vect{d}_{0k}|^{2}
 =\frac{\hbar\mu_\mathrm{e}'^{2}}{4\pi m\omega_\mathrm{e}}
 \,f_{\mathrm{rot}}
\end{equation}
where $f_{\mathrm{rot}}=1$ for the $\Sigma$ molecules,
while for molecules with a $^{2}\Pi_{3/2}$ ground state,
\begin{equation}
\label{eq3.26i5}
f_{\mathrm{rot}} = {\textstyle\frac{14}{15}}c_+^2(3/2)
 +{\textstyle\frac{2}{5}}c_-^2(3/2)
 +{\textstyle\frac{4}{15}}c_+^2(3/2)c_-^2(3/2).
\end{equation}

\begin{table}
\begin{tabular}{cccc}
\hline
&&\multicolumn{2}{c}{$\tau^{(0)}(\mathrm{s})$}\\
Species&$\frac{\omega_{k0}}{2\pi}(\mathrm{THz})$
&$T=293\mathrm{K}$&$T=77\mathrm{K}$\\
\vspace{-5 pt}\\
\hline
LiH&$\;42.1\;$&$\;25\;$
&$\;6.5\!\times\!10^9\;$\\
NH&$\;98.4\;$&$\;310,\!000\;$
&$\;1.3\!\times\!10^{25}\;$\\
OH&$\;112\;$&$\;9.8\!\times\!10^6\;$
&$\;2.2\!\times\!10^{29}\;$\\
OD&$\;81.6\;$&$\;200,\!000\;$
&$\;3.7\!\times\!10^{21}\;$\\
CaF&$\;18.4\;$&$4.7$
&$23,\!000$\\
BaF&$\;14.1\;$&$\;1.8\;$
&$\;1,\!300\;$\\
YbF&$\;15.2\;$&$\;4.1\;$
&$\;4,\!700\;$\\
LiRb&$\;5.55\;$&$\;128\;$
&$\;2,\!700\;$\\
NaRb&$\;3.21\;$&$\;1,\!400\;$
&$\;13,\!000\;$\\
KRb&$\;2.26\;$&$\;120,\!000\;$
&$\;850,\!000\;$\\
LiCs&$\;4.92\;$&$\;80\;$
&$\;1,\!300\;$\\
NaCs&$\;2.94\;$&$\;580\;$
&$\;4,\!900\;$\\
KCs&$\;1.98\;$&$\;12,\!000\;$
&$\;74,\!000\;$\\
RbCs&$\;1.48\;$&$\;63,\!000\;$
&$\;350,\!000\;$\\
\end{tabular}
\caption{
\label{Tab2b}
Lifetime against free-space vibrational heating out of the ground
state for various polar molecules at $293\,\mathrm{K}$ and
$77\,\mathrm{K}$. For comparison with the results of \cite{0772}, see
\cite{endnote}.
}
\end{table}
The calculated lifetimes for free-space vibrational heating out of the
ground state are given in Tab.~\ref{Tab2b} for $T=293\,\mathrm{K}$ and
$T=77\,\mathrm{K}$. These lifetimes are mainly determined by the
vibrational transition frequencies. Since $\omega_\mathrm{e}$
$\!\propto$ $\!1/\sqrt{m}$, the lightest molecules have the highest
vibration frequencies, which lie above the 17\,THz peak of the room
temperature spectrum, whilst the heaviest molecules vibrate well below
this frequency.  The vibrational transition frequencies of CaF, BaF
and YbF fall close to this maximum, and of the molecules considered
these three also have the largest values of $\mu_\mathrm{e}'$. For
both reasons, the ground-state lifetimes of these molecules are
limited by vibrational heating to less than 5\,s. For LiH and LiCs the
vibrational heating is an order of magnitude slower, whilst it is
exceedingly slow for all the other molecules. This slowness is mainly
due to inefficient coupling with the thermal radiation which occurs
both for the heavy molecules LiRb, NaRb, KRb, NaCs, KCs and RbCs
whose vibration frequencies are too low and, even more strikingly, for
the light molecules NH, OH and OD whose frequencies are too high.
\begin{figure}[!t!]
\includegraphics[width=\linewidth]{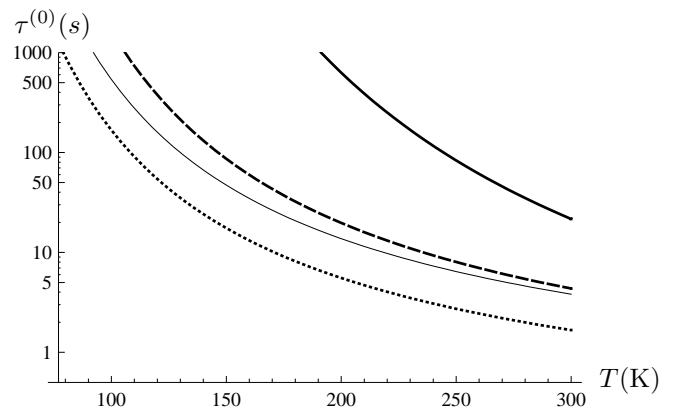}
\caption{
\label{Fig1b}
Ground state vibrational heating lifetimes in free space \textit{vs}
temperature for LiH (thick solid line), CaF (thick dashed line), YbF
(thin solid line), and BaF (thick dotted line).}
\end{figure}%
Due to the large transition frequencies, the impact of lowering the
environment temperature is even more striking for vibrational heating
than in the rotational case. This is illustrated in Fig.~\ref{Fig1b}
where the temperature-dependence of the lifetimes is displayed for the
molecules LiH, CaF, BaF and YbF which are most strongly affected by
vibrational heating.

The relative importance of rotational \textit{vs} vibrational heating
varies from molecule to molecule. Rotational heating dominates for the
hydrides and for NaCs and KCs, while vibrational heating is dominant
for the fluorides in the list, and for the other alkali dimers.

We have confined our attention to the rates for rotational and
vibrational excitation out of the ground state. The calculation is
very easily adapted to the excited states, remembering that then there
will be both excitation processes to higher lying states, and
de-excitation processes to lower lying ones. For the latter processes,
$n(\omega_{k0})$ should be replaced with $n(\omega_{k0})+1$ to account
for spontaneous emission. The calculations also need to be modified if
applied electric or magnetic fields are present, so as to account for
the Stark or Zeeman shifted transition frequencies, and any associated
change in the transition dipole moments.


\subsection{Molecules near a surface}
\label{Sec4.2}

We turn now to the question of how proximity to a surface can affect
the heating rate. Let us consider a molecule at distance $z_{\!A}$
from the surface of a homogeneous magneto-electric body of (relative)
permittivity $\varepsilon(\omega)$ and (relative) permeability
$\mu(\omega)$. The body can be modelled by a semi-infinite half space
provided it is close enough to the molecule and sufficiently smooth.
The scattering part of the Green tensor is then given by
\begin{multline}
\label{eq4.2.1}
\ten{G}^{(1)}(\vect{r},\vect{r},\omega)\\
=\frac{\mi}{8\pi}\int _0^\infty \dif q\, \frac{q}{\beta}\,
e^{2i\beta z}
\biggl[\biggl(r_s-\frac{\beta^2c^2}{\omega^2}\,r_p\biggr)
(\vect{e}_x\vect{e}_x+\vect{e}_y\vect{e}_y)\\
+2\,\frac{q^2c^2}{\omega^2}\,r_p\vect{e}_z\vect{e}_z\biggr]
\end{multline}
where
\begin{equation}
\label{eq4.2.2}
r_s=\frac{\mu(\omega)\beta-\beta_1}
 {\mu(\omega)\beta+\beta_1}\,,\qquad
r_p=\frac{\varepsilon(\omega)\beta-\beta_1}
 {\varepsilon(\omega)\beta+\beta_1}
\end{equation}
are the reflection coefficients for $s$- and $p$-polarised waves,
\begin{equation}
\label{eq4.2.3}
\beta=\sqrt{\frac{\omega^2}{c^2}-q^2},\qquad
\beta_1
=\sqrt{\frac{\omega^2}{c^2}\,\varepsilon(\omega)\mu(\omega)-q^2}
\end{equation}
($\operatorname{Im}\beta$, $\operatorname{Im}\beta_1$ $\!\ge$ $\!0$)
denote the $z$-component of the wave vector in free space ($\beta$) and
inside the half space ($\beta_1$) and $q$ is its component parallel to
the surface. For computational purposes, it is often convenient to
express the Green tensor as an integral over $\beta$,
\begin{multline}
\label{eq4.2.4}
\ten{G}^{(1)}(\vect{r},\vect{r},\omega)\\
=\frac{\mi}{8\pi}\int _0^{\omega/c} \dif\beta\,
e^{2i\beta z}
\biggl[\biggl(r_s-\frac{\beta^2c^2}{\omega^2}\,r_p\biggr)
(\vect{e}_x\vect{e}_x+\vect{e}_y\vect{e}_y)\\
+2\biggl(1-\frac{\beta^2c^2}{\omega^2}\biggr)
 r_p\vect{e}_z\vect{e}_z\biggr]\\
+\frac{1}{8\pi}\int _0^\infty\dif b\,
e^{-2bz}
\biggl[\biggl(r_s+\frac{b^2c^2}{\omega^2}\,r_p\biggr)
(\vect{e}_x\vect{e}_x+\vect{e}_y\vect{e}_y)\\
+2\biggl(1+\frac{b^2c^2}{\omega^2}\biggr)
 r_p\vect{e}_z\vect{e}_z\biggr]
\end{multline}
($\beta$ $\!=$ $\!\mi b$). Here, the first term represents the
oscillating contributions due to travelling waves, while the second
term contains the exponentially decaying contributions from evanescent
waves.

Transition rates for a molecule near a half space can be obtained by
substituting the scattering part of the Green tensor $\ten{G}^{(1)}$
[as given by Eq.~(\ref{eq4.2.1}) or (\ref{eq4.2.4})] together with its
free-space part [Eq.~(\ref{eq3.22})] into Eq.~(\ref{eq3.5}). The
ground-state heating rates then take the particularly simple form
\begin{eqnarray}
\label{eq4.2.4b}
\Gamma(\vect{r}_{\!A})&\!=&\!\Gamma^{(0)}+\Gamma^{(1)}(\vect{r}_{\!A})
 \nonumber\\
&\!=&\!\Gamma^{(0)}\left[1+\frac{2\pi c}{\omega_{k0}}\operatorname{Im}\trace
\ten{G}^{(1)}(\vect{r}_{\!A},\vect{r}_{\!A},\omega_{k0})\right].
\end{eqnarray}
In general, the integral appearing in Eq.~(\ref{eq4.2.1}) or
(\ref{eq4.2.4}) has to be evaluated numerically, but analytic results
can be obtained for sufficiently small or large molecule--surface
separations. The non-retarded limit applies to short distances, where
$z|\sqrt{\varepsilon\mu}|\omega/c$ $\!\ll$ $\!1$, while the retarded
limit holds for long distances such that \mbox{$z\omega/c$ $\!\gg$
$\!1$}. For the materials we consider in this paper,
$|\sqrt{\varepsilon\mu}|$ takes on values between $65$ and $27,\!000$
for rotational transitions and between $4.5$ and $700$ for vibrational
transitions, depending on the molecule and the material. Consequently,
there is quite a large range of intermediate distances where neither
limit applies.

In the non-retarded limit, the Green tensor~(\ref{eq4.2.4}) is
dominated by the integral over evanescent waves, which effectively
extends up to a wave vector $b$ $\!=$ $\!1/(2z)$. Over most of this
region, $\beta$ $\!\simeq$ $\!\beta_1$ $\!\simeq\mi q$, allowing us to
use the approximations
\begin{equation}
\label{eq4.2.5}
r_s\simeq\frac{\mu(\omega)-1}
 {\mu(\omega)+1}\,,\qquad
r_p\simeq\frac{\varepsilon(\omega)-1}
 {\varepsilon(\omega)+1}\,.
\end{equation}
Performing the remaining integral and retaining only the leading order
in $z\omega/c$, one finds that in this non-retarded limit, the Green
tensor is well approximated by \cite{Henkel99}
\begin{multline}
\label{eq4.2.6}
\ten{G}^{(1)}(\vect{r},\vect{r},\omega)\\
=\frac{c^2}{32\pi\omega^2z^3}\,
 \frac{\varepsilon(\omega)-1}{\varepsilon(\omega)+1}\,
(\vect{e}_x\vect{e}_x+\vect{e}_y\vect{e}_y+2\vect{e}_z\vect{e}_z).
\end{multline}
Note that by retaining only the leading order in $z\omega/c$, the
dependence on $r_{s}$ and thus also that on $\mu$ has vanished. In any
case, $\mu$ is close to 1 even for the ferromagnetic metals at the
typical frequencies of interest here (i.e. $\omega/2\pi >
10\,\mathrm{GHz}$). On substituting Eq.~(\ref{eq4.2.6}) into
Eq.~(\ref{eq3.5}), we obtain the approximate, near-field transition
rate
\begin{multline}
\label{eq4.2.7}
\Gamma_{nk}(z_{\!A})
=\Gamma_{nk}^{(0)}+\frac{|\vect{d}_{nk}|^2+|d_{nk,z}|^2}
 {8\pi\varepsilon_0\hbar z_{\!A}^3}\,
 \frac{\operatorname{Im}\varepsilon(\omega_{nk})}
 {|\varepsilon(\omega_{nk})+1|^2}\\
\times\{\Theta(\omega_{nk})[n(\omega_{nk})+1]
 -\Theta(\omega_{kn})n(\omega_{kn})\}.
\end{multline}

In particular, the ground-state heating rates~(\ref{eq4.2.4b}) are
approximated by
\begin{equation}
\label{eq4.2.7b}
\Gamma(z_{\!A})=\Gamma^{(0)}\biggl(1
 +\frac{z_\mathrm{nr}^{3}}{z_{\!A}^3}
\biggr)
\end{equation}
where
\begin{equation}
\label{eq4.2.7d}
z_\mathrm{nr}=\frac{c}{\omega_{k0}}\,
\sqrt[3]{\frac{\operatorname{Im}\varepsilon(\omega_{k0})}
 {2|\varepsilon(\omega_{k0})+1|^2}}
\end{equation}
is a scaling length that applies to calculations in the non-retarded
limit. For a metal with permittivity
\begin{equation}
\label{eq4.2.7e}
\varepsilon(\omega)=1-\frac{\omega_\mathrm{P}^2}
 {\omega(\omega+\mi\gamma)}
\end{equation}
and provided that the transition frequency is sufficiently small,
\mbox{$\omega_{k0}\ll\gamma\le\omega_\mathrm{P}$},
$z_\mathrm{nr}$ may be estimated by the simple relation
\begin{equation}
\label{eq4.2.7f}
z_\mathrm{nr}=c\,
\sqrt[3]{\frac{\gamma}{2\omega_\mathrm{P}^2\omega_{k0}^2}}\,.
\end{equation}
The plasma frequency, $\omega_\mathrm{P}$, and damping constant,
$\gamma$, are given for various conductors in Tab.~\ref{Tab3a}.

We stress that Eq.~(\ref{eq4.2.7b}) applies only in the non-retarded
limit, and that the distance $z_\mathrm{nr}$ typically lies well
outside this limit. We define a second relevant length scale,
$z_\mathrm{c}$, the characteristic distance at which the
surface-induced rate becomes equal to the free-space rate. This does
not coincide with $z_\mathrm{nr}$, because the non-retarded limit is
not valid at this distance. We have calculated $z_\mathrm{c}$  by
numerical integration of Eqs.~(\ref{eq4.2.4}) and (\ref{eq4.2.4b}) and
we present the results for molecules near a gold surface in
Tab.~\ref{Tab3}, and for a range of other conductors in
App.~\ref{AppB}. The table also gives the corresponding values of
$z_\mathrm{nr}$, which are typically $2$--$5$ times smaller.

Since Eq.~(\ref{eq4.2.7b}) does not apply at length scales in the
vicinity of the critical distance, we searched for an alternative
formula by fitting to the numerical results obtained from the
integration of Eqs.~(\ref{eq4.2.4}) and (\ref{eq4.2.4b}) at distances
$z$ $\!\le$ $\!z_\mathrm{c}$. We find that for the molecules and
surface materials studied, the heating rates throughout this range are
well approximated by the empirical formula
\begin{equation}
\label{Mike}
\Gamma(z_{\!A})=\Gamma^{(0)}\biggl(1+\frac{z_\mathrm{c}^2}{z_{\!A}^2}
 +\frac{z_\mathrm{nr}^{3}}{z_{\!A}^3}\biggr).
\end{equation}
Furthermore, a fit to the set of critical distances for rotational
heating given in App.~\ref{AppB}, suggests the approximate formula
\begin{equation}
\label{Mike2}
z_\mathrm{c}\simeq\frac{3c}{4}\,
\sqrt[4]{\frac{\gamma}{2\omega_\mathrm{P}^2\omega_{k0}^3}}\,.
\end{equation}
This empirical formula was found to be accurate to within 1\% for all
the surfaces and molecules considered, except in cases where the
critical distances are particularly small (the hydrides and KCs),
where deviations between 1\% and 10\% are more typical. The same
formula does not accurately predict the critical distances for
vibrational heating, but these are of less importance due to their
very small values. We stress again that Eq.~(\ref{Mike}) is only
empirical as the $z_{\!A}^{-2}$ term has no physical interpretation.

\begin{table}
\begin{tabular}{cccc}
\hline
Material&$\omega_\mathrm{P}(\mbox{rad/s})$
&$\gamma(\mbox{rad/s})$
&$\omega_\mathrm{P}^2/\gamma(\mbox{rad/s})$\\
\hline
Au&$\;1.37\!\times\!10^{16}\;$
&$\;4.12\!\times\!10^{13}\;$&$\;4.53\!\times\!10^{18}\;$\\
Al&$\;2.25\!\times\!10^{16}\;$
&$\;1.22\!\times\!10^{14}\;$&$\;4.15\!\times\!10^{18}\;$\\
Pd&$\;8.36\!\times\!10^{15}\;$
&$\;2.16\!\times\!10^{13}\;$&$\;3.24\!\times\!10^{18}\;$\\
Ag&$\;5.77\!\times\!10^{15}\;$
&$\;1.15\!\times\!10^{13}\;$&$\;2.89\!\times\!10^{18}\;$\\
Cu&$\;1.12\!\times\!10^{16}\;$
&$\;4.41\!\times\!10^{13}\;$&$\;2.87\!\times\!10^{18}\;$\\
Mo&$\;1.14\!\times\!10^{16}\;$
&$\;7.86\!\times\!10^{13}\;$&$\;1.65\!\times\!10^{18}\;$\\
Fe&$\;6.23\!\times\!10^{15}\;$
&$\;2.79\!\times\!10^{13}\;$&$\;1.39\!\times\!10^{18}\;$\\
Co&$\;1.18\!\times\!10^{16}\;$
&$\;1.07\!\times\!10^{14}\;$&$\;1.29\!\times\!10^{18}\;$\\
W&$\;9.72\!\times\!10^{15}\;$
&$\;8.53\!\times\!10^{13}\;$&$\;1.11\!\times\!10^{18}\;$\\
Ni&$\;7.44\!\times\!10^{15}\;$
&$\;6.53\!\times\!10^{13}\;$&$\;8.49\!\times\!10^{17}\;$\\
Pt&$\;7.75\!\times\!10^{15}\;$
&$\;1.04\!\times\!10^{14}\;$&$\;5.75\!\times\!10^{17}\;$\\
ITO&$\;3.33\!\times\!10^{15}\;$
&$\;1.68\!\times\!10^{14}\;$&$\;6.63\!\times\!10^{16}\;$\\
\end{tabular}
\caption{
\label{Tab3a}
Drude parameters for various conductors. Values are taken
from Ref.~\cite{0814}, with the exception of those for ITO (indium tin
oxide) \cite{0815}. The list is in order of decreasing
$\omega_\mathrm{P}^2/\gamma$, which corresponds to increasing surface
heating rate.
}
\end{table}
\begin{table}
\begin{tabular}{ccccc}
\hline
&\multicolumn{2}{c}{Rotational}&\multicolumn{2}{c}{\;Vibrational}\\
\;Species\;&\;$z_\mathrm{nr}(\mu\mathrm{m})$\;
&\;$z_\mathrm{c}(\mu\mathrm{m})$\;
&\;$z_\mathrm{nr}(\mu\mathrm{m})$\;
&\;$z_\mathrm{c}(\mu\mathrm{m})$\;\\
\hline
LiH&$0.73$&$1.9$&$0.035$&$0.071$\\
NH&$0.42$&$1.0$&$0.020$&$0.042$\\
OH&&&$0.018$&$0.0039$\\
(a)&$0.23$&$0.50$\\
(b)&$0.17$&$0.36$\\
(c)&$0.13$&$0.27$\\
(d)&$0.10$&$0.20$\\
OD&&&$0.022$&$0.0048$\\
(a)&$0.34$&$0.78$\\
(b)&$0.17$&$0.35$\\
(c)&$0.15$&$0.30$\\
(d)&$0.12$&$0.24$\\
CaF&$5.5$&$19$&$0.061$&$0.12$\\
BaF&$7.8$&$27$&$0.072$&$0.14$\\
YbF&$7.1$&$25$&$0.069$&$0.14$\\
LiRb&$7.6$&$26$&$0.13$&$0.27$\\
NaRb&$17$&$64$&$0.19$&$0.41$\\
KRb&$24$&$98$&$0.24$&$0.54$\\
LiCs&$8.2$&$29$&$0.15$&$0.30$\\
NaCs&$3.9$&$13$&$0.21$&$0.44$\\
KCs&$1.3$&$3.6$&$0.27$&$0.60$\\
RbCs&$42$&$180$&$0.33$&$0.75$\\
\end{tabular}
\caption{
\label{Tab3}
Non-retarded length scales and critical distances for surface
enhancement of rotational and vibrational heating rates near a gold
surface.}
\end{table}

The critical distances given in Tab.~\ref{Tab3} show that the surface
does not generate any significant heating when the molecules are more
than a few hundred $\mu\mathrm{m}$ away. However, if the molecules are
held a few $\mu\mathrm{m}$ from a surface, as they might be on a
molecule chip, there is a substantial increase in the rotational
heating for all the molecules considered, apart from the hydrides.
Even in cases where the free-space rate is small, the enhanced rate
can be very large because of the rapid inverse-power scaling. For
example, in free space, the rotational heating time of CaF, 
$3400\,\mathrm{s}$, is enormous compared with the vibrational lifetime
of $4.7\,\mathrm{s}$. However, at a distance of $1\,\mu\mathrm{m}$
from a room temperature gold surface the lifetime for rotational
excitation drops to about 8\,s and at smaller distances the rotational
heating rate dominates over the vibrational rate. For the hydrides,
the high rotational frequency that gives them  rapid free-space
heating also makes them relatively insensitive to the proximity of the
surface except at sub-micron distances.

\begin{figure}[!t!]
\includegraphics[width=\linewidth]{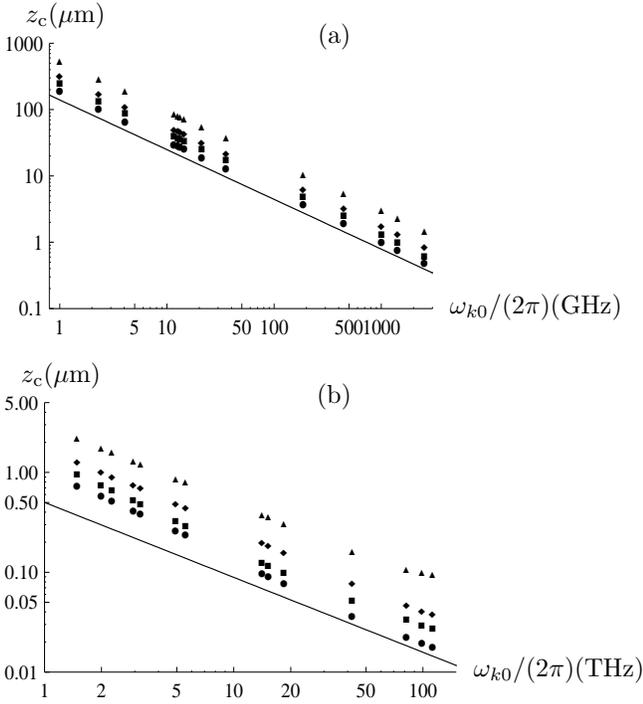}
\caption{
\label{Fig3b}
Exact critical distances for surface-induced heating \textit{vs}
frequency of the molecular transition. (a) Rotational heating. (b)
Vibrational heating. Surface materials are gold (circles), iron (squares),
platinum (diamonds) and ITO (triangles). Frequencies of the plotted data
points correspond (left to right) to RbCs, KRb, NaRb, LiCs, BaF, LiRb,
YbF, CaF, NaCs, KCs, LiH, NH, OD(a) and OH(a). Solid lines indicate
the slope corresponding to the $\omega^{-3/4}$ frequency dependence
given by the empirical formula~(\ref{Mike2}).}
\end{figure}

Figure~\ref{Fig3b}(a) shows the critical distances for rotational
heating of various molecules near a range of surfaces. It is seen
that their frequency scaling follows quite nicely the $\omega^{-3/4}$
dependence given by Eq.~(\ref{Mike2}), which is indicated by the
solid line. This trend continues in Fig.~\ref{Fig3b}(b), which shows
the critical distances for vibrational heating. These are, of
course, smaller because the vibrational frequencies are higher.

Equation~(\ref{eq4.2.7f}) shows that the short-range heating depends
on the surface material through the factor
$\omega_\mathrm{P}^2/\gamma$. A low value of this ratio leads to a
large critical distance and hence to more surface-induced heating. The
values are displayed in the last column of Tab.~\ref{Tab3a} for
various metals in order of decreasing $\omega_\mathrm{P}^2/\gamma$. In
this list, gold is the metal of choice when trying to minimise
surface-induced heating of the molecules, as also indicated by the
circles in Fig.~\ref{Fig3b}. At the opposite extreme lies indium tin
oxide (ITO), which we include here because of its attractive
combination of conductivity and optical transparency. It has a low
plasma frequency and a high damping rate $\gamma$ and so generates
stronger heating, as shown by the triangles in Fig.~\ref{Fig3b}. The
values for other metals generally obey the
$\omega_\mathrm{P}^2/\gamma$ scaling, though there are some exceptions
where values of $\omega_\mathrm{P}^2/\gamma$ are very similar.

As indicated by Eq.~(\ref{eq4.2.7d}), the surface-induced heating
will be particularly large in cases where $|\varepsilon|$ is not
large, but $\varepsilon$ has a significant imaginary part. This never
happens for a conductor, but can occur for dielectric media that
happen to be strong absorbers at the relevant frequency. As an
example, consider borosilicate glass which has $\varepsilon$ $\!=$
$\!6.2+0.21\mi$ for frequencies in the tens of GHz range
\cite{Grignon(1)2003}. The critical distance for rotational heating of
CaF near such a surface is about $620\,\mu\mathrm{m}$, very much
larger than for a metallic surface. The timescale for rotational
heating, which is $3400\,\mathrm{s}$ in free space, is thus reduced to
just $0.14\,\mathrm{s}$ when this molecule is $10\,\mu\mathrm{m}$ from
such a glass surface.

Next, we turn to the retarded limit, where \mbox{$z\omega/c$ $\!\gg$
$\!1$}, so the integrand in Eq.~(\ref{eq4.2.4}) is rapidly oscillating
or decaying over most parts of the integration regime. The main
contribution to the integral~(\ref{eq4.2.4}) comes from the region
around the stationary-phase point $q$ $\!=$ $\!0$, so that we may
approximate
\begin{equation}
\label{eq4.2.8}
r_s\simeq-r_p
\simeq\frac{\sqrt{\mu(\omega)}-\sqrt{\varepsilon(\omega)}}
 {\sqrt{\mu(\omega)}+\sqrt{\varepsilon(\omega)}}\,.
\end{equation}
The integral can then be performed, and upon retaining the leading
order in $c/(z\omega)$, one finds that the Green tensor in the
retarded limit reads
\begin{multline}
\label{eq4.2.9}
\ten{G}^{(1)}(\vect{r},\vect{r},\omega)
=\frac{e^{2\mi z\omega/c}}{8\pi z}\,
 \frac{\sqrt{\mu(\omega)}-\sqrt{\varepsilon(\omega)}}
 {\sqrt{\mu(\omega)}+\sqrt{\varepsilon(\omega)}}\\
\times(\vect{e}_x\vect{e}_x+\vect{e}_y\vect{e}_y).
\end{multline}
Consequently, the transition rates~(\ref{eq3.5}) are given by
\begin{multline}
\label{eq4.2.10}
\Gamma_{nk}(z_{\!A})
=\Gamma_{nk}^{(0)}
 +\frac{\omega_{nk}^2\bigl(|d_{nk,x}|^2\!+\!|d_{nk,y}|^2\bigr)}
 {4\pi\varepsilon_0\hbar c^2z_{\!A}}\\
 \times\operatorname{Im}\Biggl(
 \frac{\sqrt{\mu(\omega_{nk})}-\sqrt{\varepsilon(\omega_{nk})}}
 {\sqrt{\mu(\omega_{nk})}+\sqrt{\varepsilon(\omega_{nk})}}\,
 e^{2\mi z_{\!A}\omega_{nk}/c}\Biggr)\\
\times\{\Theta(\omega_{nk})[n(\omega_{nk})+1]
 -\Theta(\omega_{kn})n(\omega_{kn})\};
\end{multline}
for a good conductor, $|\varepsilon|$ $\!\gg$ $\!|\mu|$, they further
simplify to
\begin{multline}
\label{eq4.2.11}
\Gamma_{nk}(z_{\!A})
=\Gamma_{nk}^{(0)}
 -\frac{\omega_{nk}^2\bigl(|d_{nk,x}|^2\!+\!|d_{nk,y}|^2\bigr)}
 {4\pi\varepsilon_0\hbar c^2z_{\!A}}
 \sin\biggl(\frac{2z_{\!A}\omega_{nk}}{c}\biggr)\\
\times\{\Theta(\omega_{nk})[n(\omega_{nk})+1]
 -\Theta(\omega_{kn})n(\omega_{kn})\}.
\end{multline}
In particular, the ground-state heating rates~(\ref{eq4.2.4b}) are
given by
\begin{eqnarray}
\label{eq4.2.12}
\Gamma(z_{\!A})&\!=&\!\Gamma^{(0)}
 \Biggl[1+\frac{c}{2z_{\!A}\omega_{k0}}\,
 \nonumber\\
&\!&\!\times\operatorname{Im}\Biggl(
 \frac{\sqrt{\mu(\omega_{k0})}-\sqrt{\varepsilon(\omega_{k0})}}
 {\sqrt{\mu(\omega_{k0})}+\sqrt{\varepsilon(\omega_{k0})}}\,
 e^{2\mi z_{\!A}\omega_{k0}/c}\Biggr)\Biggr]\nonumber\\
&\!\simeq&\!\Gamma^{(0)}\biggl[1-\frac{c}{2z_{\!A}\omega_{k0}}\,
 \sin\biggl(\frac{2z_{\!A}\omega_{nk}}{c}\biggr)\biggr]\,.
\end{eqnarray}
Thus, the surface-induced modification of the heating rates in the
retarded limit is an oscillating function of distance, where the
amplitude of the oscillation follows a $z_{\!A}^{-1}$ power law. In
particular, the heating rates approach their free-space values in the
limit $z_{\!A}$ $\!\to$ $\!\infty$.
\begin{figure}[!t!]
\includegraphics[width=\linewidth]{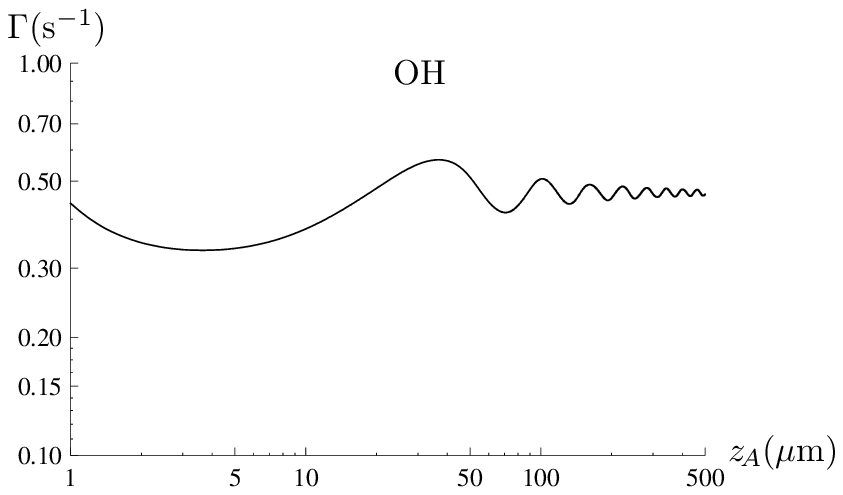}\\
\includegraphics[width=\linewidth]{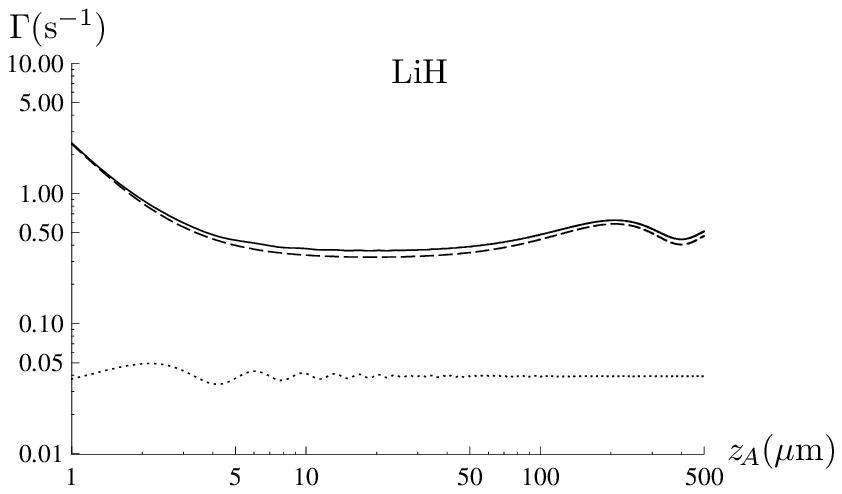}\\
\includegraphics[width=\linewidth]{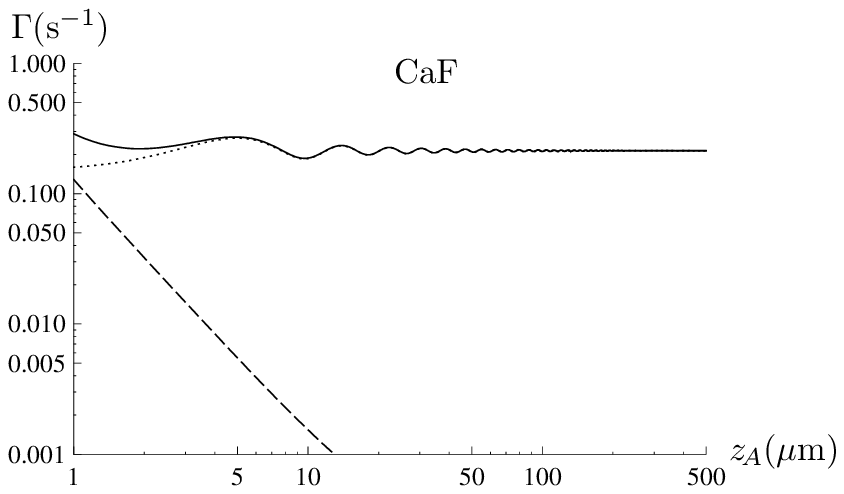}\\
\includegraphics[width=\linewidth]{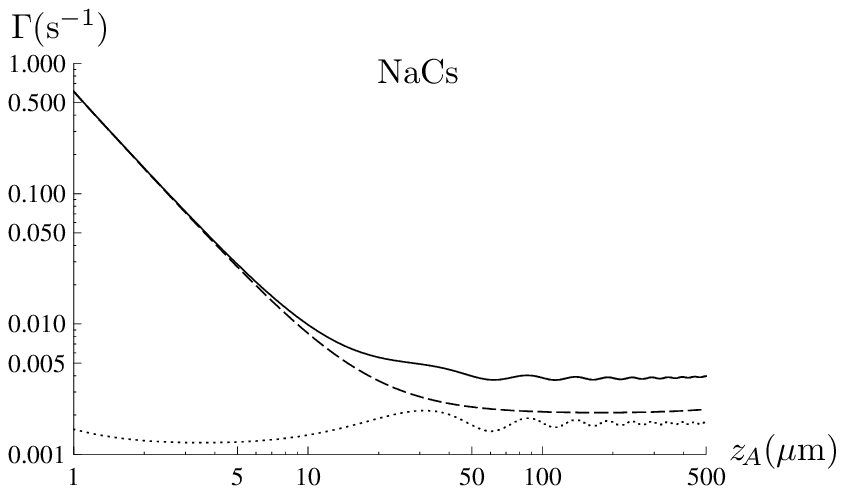}\\
\caption{
\label{Fig3}
Heating rates for OH, LiH, CaF and NaCs \textit{vs} distance from a
gold surface. Solid lines: total heating rate. Dotted lines:
vibrational excitation rate. Dashed lines: rotational excitation
rate.}
\end{figure}%

In order to see the entire distance dependence of the heating rate
it is necessary to calculate the rates as given by Eq.~(\ref{eq4.2.4})
and (\ref{eq4.2.4b}) numerically. The results are displayed in
Fig.~\ref{Fig3} where we show the total heating rates as a function of
distance for OH, LiH, CaF and NaCs molecules at distances in the range
$1-500\,\mu\mathrm{m}$ from a gold surface. In all cases, the
vibrational heating rate is dotted, the rotational rate is dashed and
the total rate is a solid line. For OH the heating is entirely
dominated by the rotational transitions over the whole of this
distance range, and the vibrational contribution does not even appear
in the plot. The heating rate is modulated with a period of
$60\,\mu\mathrm{m}$ just as expected in the retarded limit
[Eq.~(\ref{eq4.2.12})]. The heating rate is not greatly altered from
its free space value, even at the shortest distance considered. For
LiH, the heating is again dominated by the rotational transitions. The
far-field oscillations modulate the rate and we see roughly one cycle
with a period of $338\,\mu\mathrm{m}$. The heating rate rises sharply
inside the critical distance for rotational excitation, which is
$1.9\,\mu\mathrm{m}$, whereas the vibrational contribution, having a
much shorter critical distance, remains essentially constant down to
$1\,\mu\mathrm{m}$. For CaF, the heating is dominated by vibrational
excitation at $18.4\,\mathrm{THz}$, corresponding to an oscillation
period in the far field of $8\,\mu\mathrm{m}$, which can clearly be
seen. Inside the $19\,\mu\mathrm{m}$ critical distance for rotational
heating, we see a dramatic increase in the rotational contribution to
the rate, such that the two contributions are roughly equal at a
distance of $1\,\mu\mathrm{m}$ from the surface. For NaCs, the two
contributions are roughly equal in the far field and both are rather
small. The $3\,\mathrm{THz}$ vibrational heating exhibits the expected
far-field oscillations, whilst the rotational heating is at too low a
frequency to show oscillations over this range. Inside the
$13\mu\mathrm{m}$ critical distance, the rotational heating increases
rapidly, becoming a thousand times faster at a distance of
$1\,\mu\mathrm{m}$.

\begin{figure}[!t!]
\includegraphics[width=\linewidth]{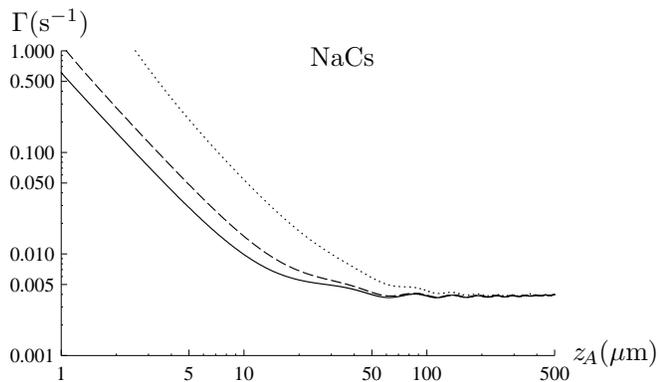}\caption{
\label{Fig4}
Heating rate for ground state NaCs as a function of
the distance from gold (solid line), iron (dashed line) and ITO (dotted
line) surfaces.
}
\end{figure}

In Fig.~\ref{Fig4}, we show once again the heating rate for NaCs as a
function of distance from a gold surface (solid line). This figure
also shows for comparison the heating rates near iron and ITO
surfaces. At distances large enough for the retarded limit to apply,
the heating rate given by Eq.~(\ref{eq4.2.12}) is independent of the
particular metallic surface since these are all good conductors at the
relevant excitation frequencies. At short range however, where the
near-field limit of Eq.~(\ref{eq4.2.7b}) applies, the heating rate
becomes proportional to $\gamma/\omega_\mathrm{P}^2$. As shown in
Tab.~\ref{Tab3a}, this ratio differs widely between these materials
and is a hundred times larger for ITO than for gold. For this reason,
the ITO surface produces a larger heating rate at short distance and
exhibits a longer critical distance than gold, as seen in
Fig.~\ref{Fig4}.

So far, we have discussed surface-enhanced heating in the presence of
metallic and dielectric surfaces. It is also interesting to consider
the heating rate for molecules in the vicinity of meta-materials,
since these offer tunable magneto-electric properties
\cite{0477,0479}, and can even be left-handed \cite{0476}. A
left-handed medium is realized when the real parts of both
$\varepsilon$ and $\mu$ are simultaneously negative, leading to a
negative index of refraction and a number of counterintuitive effects
associated with the propagation of the electromagnetic field inside
such a medium \cite{0476}. Since the surface-enhanced heating rate of
a single interface depends solely on the reflected electromagnetic
field, one would expect it to be insensitive to left-handedness. For
weakly absorbing media, the oscillations seen in the retarded limit
are small when the signs of $\operatorname{Re}\varepsilon$ and
$\operatorname{Re}\mu$ are both positive or both negative. In the case
where $\operatorname{Re}\varepsilon$ $\!=$ $\!\operatorname{Re}\mu$,
and the imaginary parts are small, there are no oscillations at all
since $r_{s}$ and $r_{p}$ are then very close to zero. The amplitude
of the oscillations is greatest when these reflection coefficients
have their maximum values of 1. As we have already seen, this occurs
for metals because $|\varepsilon|$ is much larger than $|\mu|$. In the
context of meta-materials, reflection coefficients close to unity are
obtained for any weakly absorbing medium where
$\operatorname{Re}\varepsilon$ and $\operatorname{Re}\mu$ have
opposite signs. Note that this result is insensitive to the magnitudes
of $\operatorname{Re}\varepsilon, \operatorname{Re}\mu$, which neither
need to be equal nor particularly large; they need only be of opposite
sign and considerably larger than the imaginary parts. A meta-material
engineered with these properties would produce large oscillations in
the heating rate with a phase determined by the chosen values of
$\operatorname{Re}\varepsilon, \operatorname{Re}\mu$.
\begin{figure}[!t!]
\includegraphics[width=\linewidth]{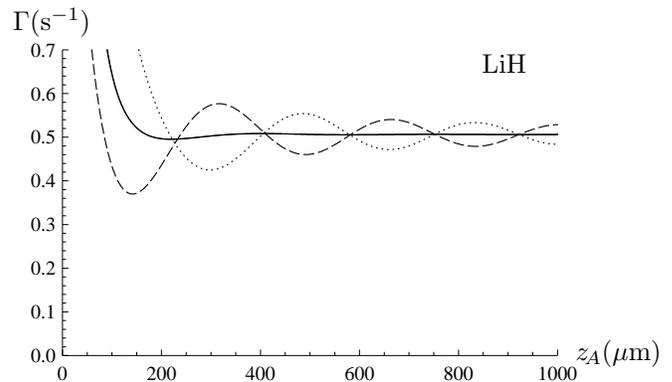}
\caption{
\label{Fig5}
Heating rate for ground state LiH as a function of distance from
fictitious meta-materials with
$\varepsilon(\omega_{k0})$ $\!=$ $\!\mu(\omega_{k0})$ $\!=$
$\!\pm 10\!+\!\mi$ (solid line),
\mbox{$\varepsilon(\omega_{k0})$ $\!=$ $\!10\!+\!\mi$},
\mbox{$\mu(\omega_{k0})$ $\!=$ $\!-10\!+\!\mi$} (dashed line) and
\mbox{$\varepsilon(\omega_{k0})$ $\!=$ $\!-10\!+\!\mi$},
\mbox{$\mu(\omega_{k0})$ $\!=$ $\!10\!+\!\mi$} (dotted line).
}
\end{figure}%
Figure~\ref{Fig5} shows the heating rate of a LiH molecule near
fictitious weakly absorbing meta-materials with
$\operatorname{Re}\varepsilon$ $\!=$ $\!\pm\operatorname{Re}\mu$.
We see that the left-handed material
(\mbox{$\operatorname{Re}\varepsilon$, $\operatorname{Re}\mu$ $\!<$
$\!0$}) gives rise to exactly the same heating rate as a comparable
ordinary material with \mbox{$\operatorname{Re}\varepsilon$,
$\operatorname{Re}\mu$ $\!>$ $\!0$} (the two curves cannot be
distinguished on the plot), and that the oscillations are suppressed.
On the other hand, media with \mbox{$\operatorname{Re}\varepsilon$
$\!=$ $-\operatorname{Re}\mu$} result in large oscillations of the
heating rate in the long-distance regime, with a phase that depends on
the material properties.


\subsection{Surfaces of finite thickness}
\label{Sec4.3}

The results of the previous section have shown that metallic surfaces
can considerably enhance surface-indu\-ced heating. In the context of
chips, metal surfaces are often unavoidable since they are used in
current- or charge-carrying structures. One possible strategy to
reduce the associated molecular heating is to reduce the thickness of
the metal substrates. For a slab of finite thickness $d$, coated onto
an infinitely thick substrate of permittivity
$\varepsilon_\mathrm{s}$ $\!=$ $\!\varepsilon_\mathrm{s}(\omega)$ and
permeability $\mu_\mathrm{s}$ $\!=$ $\!\mu_\mathrm{s}(\omega)$, the
surface-induced heating rate is still given by Eq.~(\ref{eq4.2.4b})
together with Eq.~(\ref{eq4.2.4}), but the reflection coefficients are
now given by
\begin{align}
\label{eq4.3.1b}
&r_s=
 \frac{\mu^2\beta\beta_\mathrm{s}-\mu_\mathrm{s}\beta_1^2
 +\mi\mu\beta_1[\mu_\mathrm{s}\beta-\beta_\mathrm{s}]\cot(\beta_1d)}
 {\mu^2\beta\beta_\mathrm{s}+\mu_\mathrm{s}\beta_1^2
 +\mi\mu\beta_1[\mu_\mathrm{s}\beta+\beta_\mathrm{s}]\cot(\beta_1d)}
 \,,\\
\label{eq4.3.2b}
&r_p=
 \frac{\varepsilon^2\beta\beta_\mathrm{s}
 -\varepsilon_\mathrm{s}\beta_1^2
 +\mi\varepsilon\beta_1
 [\varepsilon_\mathrm{s}\beta-\beta_\mathrm{s}]
 \cot(\beta_1d)}
 {\varepsilon^2\beta\beta_\mathrm{s}
 +\varepsilon_\mathrm{s}\beta_1^2
 +\mi\varepsilon\beta_1
 [\varepsilon_\mathrm{s}\beta+\beta_\mathrm{s}]
 \cot(\beta_1d)}
\end{align}
where
\begin{equation}
\label{eq4.2.3b}
\beta_\mathrm{s}
=\sqrt{\frac{\omega^2}{c^2}\,\varepsilon_\mathrm{s}(\omega)
 \mu_\mathrm{s}(\omega)-q^2}
\end{equation}
($\operatorname{Im}\beta_\mathrm{s}$ $\!\ge$ $\!0$).

Let us first consider the influence of the metal surface alone by
letting $\varepsilon_\mathrm{s}=\mu_\mathrm{s}=1$. In the non-retarded
limit, the reflection coefficients may then be approximated by
\begin{eqnarray}
\label{eq4.3.3}
r_s&\!\simeq&\!\frac{\mu^2-1}
 {\mu^2+1+2\mu\coth(qd)}\,,\qquad\\
\label{eq4.3.4}
r_p&\!\simeq&\!\frac{\varepsilon^2-1}
 {\varepsilon^2+1+2\varepsilon\coth(qd)}
\end{eqnarray}
[recall the discussion above Eq.~(\ref{eq4.2.5})]. Since $q$
$\!\lesssim$ $\!1/(2z_{\!A})$, the short-range heating rate will be
identical to that of an infinitely thick plate provided $d$ $\!\gg$
$\!z_{\!A}$, since in this limit the above reflection coefficients
reduce to those given in Eqs.~(\ref{eq4.2.2}). On the other hand, the
reflection coefficients, and hence also the heating rate, must become
very small when $d|\varepsilon|$ $\!\ll$ $\!z_{\!A}$. We note
immediately that, for molecule--surface separations of interest, a
conducting surface needs to be unfeasibly thin for this limit to be
reached, because of the enormously large values of $|\varepsilon|$ for
a conductor. The behaviour between the two limits has to be determined
from a numerical analysis.

In the retarded limit, one may approximate
\begin{equation}
\label{eq4.3.5}
r_s\simeq-r_p\simeq
\frac{\varepsilon-\mu}{\varepsilon+\mu
 +2\mi\sqrt{\varepsilon\mu}\,
 \cot\bigl(\sqrt{\varepsilon\mu}\,\omega d/c\bigr)}
\end{equation}
so for a good conductor, $|\varepsilon|$ $\!\gg$ $\!|\mu|$, the
reflection coefficients and the heating rate become independent of the
plate thickness at long range.

\begin{figure}[!t!]
\includegraphics[width=\linewidth]{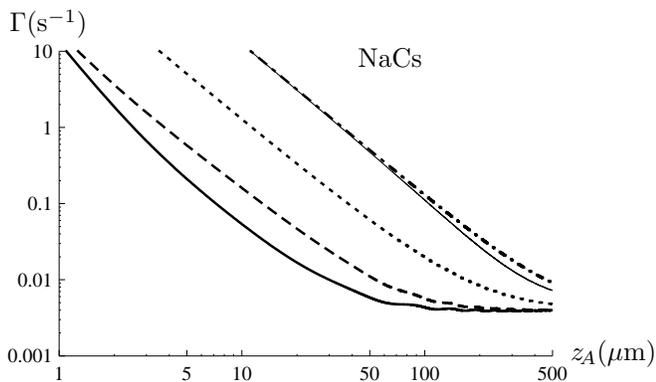}\\[3ex]
\caption{
\label{Fig6}
Heating rate of NaCs as a function of distance from an ITO surface
with thickness $10\,\mu\mathrm{m}$ (thick solid line),
$1\,\mu\mathrm{m}$ (dashed line), $0.1\,\mu\mathrm{m}$ (dotted line)
and $0.01\,\mu\mathrm{m}$ (dashed-dotted line). The thin solid line
shows how the result changes for a $0.01\,\mu\mathrm{m}$ thick surface
when the ITO is coated onto borosilicate glass.}
\end{figure}

In Fig.~\ref{Fig6}, we display the surface-induced heating rate of a
NaCs molecule near ITO plates of various thicknesses as a function of
molecule--plate separation.
At large separations, the heating rates are independent of the
thickness, as predicted from Eq.~(\ref{eq4.3.5}). Over the entire
range of distances calculated, the ITO plate of thickness
$\!10\mu\mathrm{m}$ gives the same result as a plate of infinite
thickness. Reducing the thickness below this value increases the
heating rates at short distances, contrary to the expectation of
reduced rates at short range anticipated from Eqs.~(\ref{eq4.3.3}) and
(\ref{eq4.3.4}). A reduction of the short-range heating rates below
the values observed for thick plates is eventually found but only once
the coating is unfeasibly thin. Thus our calculations show that the
heating at short-range cannot be reduced by reducing the material
thickness. To understand the increase of the heating rate with
decreasing thickness, note that in the non-retarded limit the heating
is mainly due to the coupling of the molecule with the surface
plasmons at the front face of the plate. As the thickness decreases,
these couple to the plasmons at the back face of the plate,
leading to mutual enhancement and thus to an increase of the heating
rate \cite{Lenac}.
To include the borosilicate glass substrate we took
$\varepsilon_\mathrm{s}(\omega_{k0})$ $\!=$ $\!6.2+0.21\mi$ for the
rotational transitions \cite{Grignon(1)2003} and
$\varepsilon_\mathrm{s}(\omega_{k0})$ $\!=$ $\!6.4+0.74\mi$ for the
vibrational ones \cite{Naftaly07}. We find identical results  whether
or not this substrate is included, except for the thinnest coating,
$d=0.01\,\mu$m, where we find that the presence of the substrate
slightly reduces the heating rate, as shown by the thin solid line in
the figure.


\section{Summary and conclusions}
\label{Sec5}

Using macroscopic QED, we have calculated the internal dynamics of a
molecule placed within an arbitrary environment of magneto-electric
bodies of uniform temperature. The internal time evolution is governed
by the molecular transition frequencies and transition rates which
depend on position and temperature. The dependence on temperature is
due to the thermal photon number and can be completely understood from
considering the free-space case, while the position-dependence, which
enters via the classical Green tensor for the electromagnetic field in
the environment, can be derived from the behaviour at zero
temperature.

We have used the general formulae to study the ground-state heating
rates of several polar molecules of current experimental interest, as
a function of the distance from various surfaces. We have given a
simple approximate formula that can be used to estimate the heating
rates for any other molecules at any distance from any surface of
interest. For light molecules, particularly the hydrides, rotational
heating dominates and limits the free-space lifetime of the ground
state to a few seconds when the environment is at room temperature.
For the metal fluorides we studied, vibrational heating dominates and
again the room temperature free-space lifetime of the ground state is
of the order of a few seconds. When the molecules approach a metallic
surface, the heating rate can be greatly enhanced. This is
particularly true for the rotational transitions where the critical
distance at which the surface dominates the free-space rate is
typically in the $1-100\,\mu\mathrm{m}$ range. For the hydrides, the
free-space heating rate is large because the rotational frequencies
are large, but this same fact also means that the critical distance
for surface-induced heating is rather small. Therefore, these
molecules could be trapped up to a few microns from a surface with
little change in the heating rate. The effect of the surface on
rotational heating is very much stronger for the heavier molecules,
but since the rate in free-space is typically very small for these
molecules, they too have lifetimes of a second or more at distances up
to 1\,$\mu$m from the surface.

We have shown that, at long range, the heating rates become
insensitive to the particular surface properties, while at short-range
the heating is faster for smaller values of the parameter
$\omega_\mathrm{P}^2/\gamma$. Of the metals considered, gold induces
the lowest heating rate. We have also shown that decreasing the
thickness of the surface tends to increase the heating rate,
particularly at short distances. Dielectric materials that are good
absorbers at the relevant frequency result in large critical distances
and hence very large heating rates at short range.

In the context of
molecule chips, where confinement of molecules a few microns from the
chip surface is envisaged, our work shows that surface-induced heating
should be considered carefully when selecting appropriate molecules
and surfaces, but that confinement for several seconds in single
quantum-states is quite feasible even when the surface is at room
temperature. In all cases, lowering the environment temperature will
allow for even longer lifetimes. For approach distances smaller than
$1\,\mu\mathrm{m}$, surface-induced heating becomes rapidly
problematic, and cooling to cryogenic temperatures may be required.


\acknowledgments

This work was supported by the Alexander von Humboldt Foundation, the
Royal Society, and the UK Engineering and Physical Sciences Research
Council. The research leading to these results has received funding
from the European Community's Seventh Framework Programme
FP7/2007-2013 under grant agreement 216774. S.Y.B is grateful to
W.L.~Barnes, S.~Franzen, C.~Henkel, J.~Kirkpatrick, G.J.~McPhee,
B.E.~Sernelius and M.S.~Toma\v{s} for dicussions.


\appendix
\section{Markov approximation}
\label{AppA}

Substituting the formal solution
\begin{multline}
\label{A1}
\hat{\vect{f}}_\lambda(\vect{r},\omega,t)
=\me^{-\mi\omega t}\hat{\vect{f}}_\lambda(\vect{r},\omega)\\
+\frac{\mi}{\hbar}\sum_{m,n}\int_0^t\dif\tau\,\me^{-\mi\omega(t-\tau)}
 \vect{d}_{mn}\sprod\ten{G}_\lambda^\ast
 (\vect{r}_{\!A},\vect{r},\omega)\hat{A}_{mn}(\tau)
\end{multline}
to Eq.~(\ref{eq3.1b}) into Eq.~(\ref{eq3.1}) and using the integral
relation~(\ref{eq2.12}), one obtains
\begin{multline}
\label{A2}
\dot{\hat{A}}_{mn}(t)
 =\mi\omega_{mn} \hat{A}_{mn}(t)
 +\frac{\mi}{\hbar}\sum_k\int_0^\infty\dif\omega\\
 \times\biggl\{\me^{-\mi\omega t}
 \bigl[\vect{d}_{nk}\hat{A}_{mk}(t)-\vect{d}_{km}\hat{A}_{kn}(t)
 \bigr]\sprod
 \underline{\hat{\vect{E}}}(\vect{r}_{\!A},\omega)\\
+\me^{\mi\omega t}
 \underline{\hat{\vect{E}}}{}^\dagger(\vect{r}_{\!A},\omega)\sprod
 \bigl[\vect{d}_{nk}\hat{A}_{mk}(t)-\vect{d}_{km}\hat{A}_{kn}(t)
 \bigr]\biggr\}\\
+\hat{Z}_{mn}(t)
\end{multline}
with
\begin{multline}
\label{A3}
\hat{Z}_{mn}(t)
=-\frac{\mu_0}{\hbar\pi}
 \sum_{k,l,j}\int_0^\infty\dif\omega\omega^2
 \int_0^t\dif\tau\\
\times\bigl\{
 \bigl[\me^{-\mi\omega(t-\tau)}\hat{A}_{mk}(t)\hat{A}_{lj}(\tau)
 -\me^{\mi\omega(t-\tau)}\hat{A}_{lj}(\tau)\hat{A}_{mk}(t)\bigr]\\
\times\vect{d}_{nk}\sprod\operatorname{Im}
 \ten{G}(\vect{r}_{\!A},\vect{r}_{\!A},\omega)\sprod\vect{d}_{lj}\\
-\bigl[\me^{-\mi\omega(t-\tau)}\hat{A}_{kn}(t)\hat{A}_{lj}(\tau)
 -\me^{\mi\omega(t-\tau)}\hat{A}_{lj}(\tau)\hat{A}_{nk}(t)\bigr]\\
\times\vect{d}_{km}\sprod\operatorname{Im}
 \ten{G}(\vect{r}_{\!A},\vect{r}_{\!A},\omega)\sprod\vect{d}_{lj}
 \bigr\}
\end{multline}
denoting the zero-point contribution to the internal molecular
dynamics. This differential equation can be solved iteratively by
substituting the self-consistent solution
\begin{multline}
\label{A4}
\hat{A}_{mn}(t)=\me^{\mi\tilde{\omega}_{mn}t}\hat{A}_{mn}(0)\\
 +\frac{\mi}{\hbar}\sum_k\int_0^\infty\dif\omega
 \int_0^t\dif\tau\,\me^{\mi\tilde{\omega}_{mn}(t-\tau)}\\
 \times\biggl\{\me^{-\mi\omega\tau}
 \bigl[\vect{d}_{nk}\hat{A}_{mk}(\tau)
 -\vect{d}_{km}\hat{A}_{kn}(\tau)\bigr]\sprod
 \underline{\hat{\vect{E}}}(\vect{r}_{\!A},\omega)\\
+\me^{\mi\omega\tau}
 \underline{\hat{\vect{E}}}{}^\dagger(\vect{r}_{\!A},\omega)\sprod
 \bigl[\vect{d}_{nk}\hat{A}_{mk}(\tau)
 -\vect{d}_{km}\hat{A}_{kn}(\tau)\bigr]\biggr\}
\end{multline}
to the truncated Eq.~(\ref{A2}) without $\hat{Z}_{mn}(t)$ back into
Eq.~(\ref{A2}), where at this level of approximation, the operator
ordering in Eq.~(\ref{A4}) may be chosen arbitrarily. Taking
expectation values according to Eqs.~(\ref{eq2.14})--(\ref{eq2.17}),
one arrives at
\begin{equation}
\label{A5}
\bigl\langle\dot{\hat{A}}_{mn}(t)\bigl\rangle
 =\mi\omega_{mn}\bigl\langle\hat{A}_{mn}(t)\bigl\rangle
 +\bigl\langle\hat{T}_{mn}(t)\bigl\rangle
 +\bigl\langle\hat{Z}_{mn}(t)\bigl\rangle
\end{equation}
where
\begin{multline}
\label{A6}
\bigl\langle\hat{T}_{mn}(t)\bigl\rangle\\
=-\frac{\mu_0}{\hbar\pi}
 \sum_{k,l}\int_0^\infty\dif\omega\,\omega^2n(\omega)\int_0^t\dif\tau
 \bigl[\me^{-\mi\omega(t-\tau)}+\me^{\mi\omega(t-\tau)}\bigr]\\
\times\Bigl\{\me^{\mi\tilde{\omega}_{mk}(t-\tau)}
 \bigl[\bigl\langle\hat{A}_{ml}(\tau)\bigl\rangle
 \vect{d}_{nk}\sprod\operatorname{Im}
 \ten{G}(\vect{r}_{\!A},\vect{r}_{\!A},\omega)\sprod\vect{d}_{kl}\\
-\bigl\langle\hat{A}_{lk}(\tau)\bigl\rangle
 \vect{d}_{nk}\sprod\operatorname{Im}
 \ten{G}(\vect{r}_{\!A},\vect{r}_{\!A},\omega)\sprod\vect{d}_{lm}
 \bigr]\\
-\me^{\mi\tilde{\omega}_{kn}(t-\tau)}
 \bigl[\bigl\langle\hat{A}_{kl}(\tau)\bigl\rangle
 \vect{d}_{km}\sprod\operatorname{Im}
 \ten{G}(\vect{r}_{\!A},\vect{r}_{\!A},\omega)\sprod\vect{d}_{nl}\\
-\bigl\langle\hat{A}_{ln}(\tau)\bigl\rangle
 \vect{d}_{km}\sprod\operatorname{Im}
 \ten{G}(\vect{r}_{\!A},\vect{r}_{\!A},\omega)\sprod\vect{d}_{lk}
 \bigr]\Bigr\}
\end{multline}
denotes the thermal contribution to the internal molecular dynamics.
For weak molecule--field coupling, the contributions
$\bigl\langle\hat{T}_{mn}(t)\bigl\rangle$ and
$\bigl\langle\hat{Z}_{mn}(t)\bigl\rangle$ may be evaluated by means
of the Markov approximation, i.e. by letting
$\bigl\langle\hat{A}_{mn}(\tau)\bigr\rangle$ $\!\simeq$
$\!\me^{-\mi\tilde{\omega}_{mn}(t-\tau)}
\bigl\langle\hat{A}_{mn}(t)\bigr\rangle$ and evaluating the remaining
time integrals according to
\begin{multline}
\label{A7}
\int_0^t\dif\tau\,
 \me^{-\mi(\omega-\tilde{\omega}_{mn})(t-\tau)}\\
 \simeq\pi\delta(\omega-\tilde{\omega}_{mn})
 +\mi\mathcal{P}\,\frac{1}{\tilde{\omega}_{mn}-\omega}\,.
\end{multline}
Assuming the system to be non-degenerate, so that off-diagonal
molecular density matrix decouple from each other as well as from the
diagonal ones, we arrive at Eqs.~(\ref{eq3.2}) and (\ref{eq3.3}),
together with Eqs.~(\ref{eq3.4})--(\ref{eq3.13}).


\section{Critical distances}
\label{AppB}

We have calculated the critical distances for surface-induced
enhancement of ground-state heating rates for various combinations of
molecules and materials on the basis of the data given in
Tabs.~\ref{Tab1} and \ref{Tab3a}. The results are summarised in
Tab.~\ref{TabB1}.

\begin{table*}
\begin{tabular}{cccccccccccccccccccccccc}
\hline
&\multicolumn{11}{c}{rotational}&\hspace*{2ex}&\multicolumn{11}{c}{
vibrational } \\
Spe.&Al&Pd&Ag&Cu&Mo&Fe&Co&W&Ni&Pt&ITO
&&Al&Pd&Ag&Cu&Mo&Fe&Co&W&Ni&Pt&ITO\\
\hline
LiH&$1.9$&$2.0$&$2.0$&$2.1$&$2.4$
&$2.5$&$2.6$&$2.7$&$2.8$&$3.1$&$5.4$
&&$.044$&$.038$&$.039$&$.042$&$.053$&$.051$&$.060$
&$.060$&$.064$&$.075$&$.16$\\
NH&$1.0$&$1.0$&$1.0$&$1.1$&$1.3$&$1.3$
&$1.4$&$1.4$&$1.5$&$1.7$&$2.9$
&&$.021$&$.022$&$.023$&$.023$&$.028$&$.029$
&$.031$&$.032$&$.035$&$.040$&$.097$\\
OH&&&&&&&&&&&
&&$.019$&$.020$&$.021$&$.021$&$.025$&$.027$
&$.028$&$.029$&$.032$&$.037$&$0.94$\\
(a)&$.51$&$.47$&$.43$&$.52$&$.62$&$.60$
&$.67$&$.69$&$.73$&$.82$&$1.4$\\
(b)&$.36$&$.32$&$.29$&$.37$&$.45$&$.41$
&$.48$&$.49$&$.52$&$.59$&$1.0$\\
(c)&$.26$&$.22$&$.20$&$.26$&$.32$&$.29$
&$.35$&$.36$&$.37$&$.43$&$.77$\\
(d)&$.18$&$.15$&$.13$&$.17$&$.22$&$.29$
&$.24$&$.25$&$.25$&$.30$&$.55$\\
OD&&&&&&&&&&&
&&$.025$&$.024$&$.026$&$.026$&$.032$&$.033$
&$.035$&$.036$&$.039$&$.046$&$.11$\\
(a)&$.79$&$.77$&$.73$&$.84$&$.98$&$.97$
&$1.1$&$1.1$&$1.2$&$1.3$&$2.2$\\
(b)&$.35$&$.31$&$.28$&$.36$&$.43$&$.40$
&$.47$&$.48$&$.50$&$.57$&$1.0$\\
(c)&$.30$&$.25$&$.23$&$.29$&$.36$&$.33$
&$.39$&$.40$&$.42$&$.48$&$.86$\\
(d)&$.24$&$.19$&$.17$&$.23$&$.28$&$.25$
&$.31$&$.32$&$.33$&$.38$&$.69$\\
CaF&$19$&$20$&$21$&$21$&$24$&$25$
&$26$&$26$&$28$&$31$&$53$
&&$.094$&$.074$&$.071$&$.086$&$.11$&$.098$
&$.13$&$.13$&$.13$&$.15$&$.30$\\
BaF&$28$&$30$&$31$&$31$&$35$&$37$
&$37$&$39$&$42$&$46$&$78$
&&$.12$&$.093$&$.088$&$.11$&$.14$&$.12$
&$.16$&$.16$&$.16$&$.20$&$.37$\\
YbF&$25$&$27$&$28$&$28$&$32$&$33$
&$34$&$35$&$38$&$41$&$71$
&&$.11$&$.087$&$.083$&$.10$&$.13$&$.12$
&$.15$&$.15$&$.15$&$.18$&$.35$\\
LiRb&$27$&$29$&$30$&$30$&$34$&$35$
&$36$&$38$&$40$&$44$&$76$
&&$.27$&$.22$&$.20$&$.26$&$.32$&$.39$
&$.36$&$.36$&$.38$&$.43$&$.78$\\
NaRb&$66$&$70$&$72$&$72$&$82$&$86$
&$88$&$91$&$97$&$110$&$180$
&&$.42$&$.37$&$.34$&$.43$&$.51$&$.48$
&$.55$&$.57$&$.60$&$.68$&$1.2$\\
KRb&$100$&$110$&$110$&$110$&$130$&$130$
&$130$&$140$&$150$&$160$&$280$
&&$.55$&$.51$&$.47$&$.57$&$.68$&$.65$
&$.73$&$.75$&$.79$&$.89$&$1.6$\\
LiCs&$30$&$32$&$33$&$33$&$37$&$39$
&$40$&$41$&$44$&$49$&$83$
&&$.30$&$.25$&$.22$&$.29$&$.36$&$.32$
&$.39$&$.40$&$.42$&$.48$&$.85$\\
NaCs&$13$&$14$&$14$&$14$&$16$&$17$
&$17$&$18$&$19$&$21$&$36$
&&$.45$&$.40$&$.37$&$.46$&$.55$&$.52$
&$.59$&$.61$&$.64$&$.73$&$1.3$\\
KCs&$3.7$&$3.9$&$4.0$&$4.0$&$4.6$&$4.8$
&$4.9$&$5.1$&$5.5$&$6.0$&$10$
&&$.61$&$.58$&$.54$&$.64$&$.75$&$.73$
&$.81$&$.83$&$.88$&$.99$&$1.7$\\
RbCs&$190$&$200$&$210$&$210$&$240$&$250$
&$250$&$260$&$280$&$310$&$530$
&&$.76$&$.74$&$.70$&$.81$&$.95$&$.94$
&$1.0$&$1.1$&$1.1$&$1.2$&$2.2$\\
\end{tabular}
\caption{
\label{TabB1}
The exact critical distances $z_\mathrm{c}(\mu\mathrm{m})$ for the
enhancements of ground-state heating rates of various polar molecules
due to the presence of surfaces of various materials.
}
\end{table*}
%




\begin{thebibliography}{70}
\expandafter\ifx\csname natexlab\endcsname\relax\def\natexlab#1{#1}\fi
\expandafter\ifx\csname bibnamefont\endcsname\relax
  \def\bibnamefont#1{#1}\fi
\expandafter\ifx\csname bibfnamefont\endcsname\relax
  \def\bibfnamefont#1{#1}\fi
\expandafter\ifx\csname citenamefont\endcsname\relax
  \def\citenamefont#1{#1}\fi
\expandafter\ifx\csname url\endcsname\relax
  \def\url#1{\texttt{#1}}\fi
\expandafter\ifx\csname urlprefix\endcsname\relax\def\urlprefix{URL }\fi
\providecommand{\bibinfo}[2]{#2}
\providecommand{\eprint}[2][]{\url{#2}}

\bibitem{Bethlem(1)99} H.~L. Bethlem, G.~Berden and G.~Meijer, Phys.
Rev. Lett. \textbf{83}, 1558 (1999).

\bibitem{Bethlem(1)00} H.~L. Bethlem, G.~Berden, F.~M.~H. Crompvoets,
R.~T. Jongma, A.~J.~A. van Roij and G.~Meijer, Nature \textbf{406},
491 (2000).

\bibitem{Meerakker(1)05} S.~Y.~T. van de Meerakker, P.~H.~M. Smeets,
N.~Vanhaecke, R.~T. Jongma and G.~Meijer, Phys. Rev. Lett.
\textbf{94} 023004 (2005).

\bibitem{Sawyer(1)07} B.~C. Sawyer, B.~L. Lev, E.~R. Hudson, B.~K.
Stuhl, M.~Lara, J.~L. Bohn and J.~Ye, Phys. Rev. Lett. \textbf{98},
253002 (2007).

\bibitem{Veldhoven(1)05} J.~van Veldhoven, H.~L. Bethlem and
G.~Meijer, Phys. Rev. Lett. \textbf{94}, 083001 (2005).

\bibitem{Rieger(1)05} T.~Rieger, T.~Junglen, S.~A. Rangwala, P.~W.~H.
Pinkse and G.~Rempe, Phys. Rev. Lett. \textbf{95}, 173002 (2005).

\bibitem{Weinstein(1)98} J.~D. Weinstein, R.~deCarvalho, T.~Guillet,
B.~Friedrich and J.~Doyle, Nature \textbf{395}, 148 (1998).

\bibitem{Sage(1)05} J.~M. Sage, S.~Sainis, T.~Bergeman and D.~DeMille,
Phys. Rev. Lett. \textbf{94}, 203001 (2005).

\bibitem{Hoekstra(1)07} S.~Hoekstra, J.~J. Gilijamse, B.~Sartakov,
N.~Vanhaecke, L.~Scharfenberg, S.~Y.~T. van de Meerakker and
G.~Meijer, Phys. Rev. Lett. \textbf{98}, 133001 (2007).

\bibitem[{\citenamefont{Vanhaecke and Dulieu}(2007)}]{0772}
\bibinfo{author}{\bibfnamefont{N.}~\bibnamefont{Vanhaecke}} \bibnamefont{and}
  \bibinfo{author}{\bibfnamefont{O.}~\bibnamefont{Dulieu}},
  \bibinfo{journal}{Mol. {P}hys.} \textbf{\bibinfo{volume}{105}},
  \bibinfo{pages}{1723} (\bibinfo{year}{2007}).

\bibitem{Fortagh(1)07} J.~Fort\'agh and C.~Zimmermann, Rev. Mod.
Phys. \textbf{79}, 235 (2007).

\bibitem{Meek(1)08} S.~A. Meek, H.~L. Bethlem, H.~Conrad and
G.~Meijer, Phys. Rev. Lett. \textbf{100}, 153003 (2008).

\bibitem{Andre(1)06} A.~Andr\'e, D. DeMille, J.~M. Doyle, M.~D.
Lukin, S.~E. Maxwell, P.~Rabl, R.~J. Schoelkopf and P.~Zoller, Nature
Physics \textbf{2}, 636 (2006).

\bibitem{Rabl(1)06} P.~Rabl, D.~DeMille, J.~M. Doyle, M.~D. Lukin,
R.~J. Schoelkopf and P.~Zoller, Phys. Rev. Lett. \textbf{97}, 033003
(2006).

\bibitem{Trupke(1)07} M.~Trupke, J.~Goldwin, B.~Darqui\'e, G.~Dutier,
S.~Eriksson, J.~Ashmore and E.~A. Hinds, Phys. Rev. Lett.
\textbf{99}, 063601 (2007).

\bibitem{Warken(1)07} F.~Warken, E.~Vetsch, D.~Meschede, M.~Sokolowski
and A.~Rauschenbeutel, Optics Express \textbf{15}, 11952 (2007).

\bibitem{Purcell46}
E.~M. Purcell, Phys. Rev. \textbf{69}, 681 (1946).

\bibitem{Agarwal75}
G.~S. Agarwal, Phys. Rev. A \textbf{12}, 1475 (1975).

\bibitem{0335}
R.~R. Chance, A.~Prock, and R.~Silbey, J. Chem. Phys. \textbf{60},
2744 (1974).

\bibitem{0336}
R.~R. Chance, A.~Prock, and R.~Silbey, J. Chem. Phys. \textbf{62},
771 (1975).

\bibitem{Yeung96}
M.~S. Yeung and T.~K. Gustafson, Phys. Rev. A \textbf{54}, 5227
(1996).

\bibitem{Scheel99}
S.~Scheel, L.~Kn\"oll, and D.-G. Welsch, Phys. Rev. A \textbf{60}, 4094
(1999).

\bibitem[{\citenamefont{Ho et~al.}(2003)\citenamefont{Ho, Buhmann,
Kn\"{o}ll,
  Welsch, Scheel, and K\"{a}stel}}]{0002}
\bibinfo{author}{\bibfnamefont{H.~T.} \bibnamefont{Dung}},
  \bibinfo{author}{\bibfnamefont{S.~Y.} \bibnamefont{Buhmann}},
  \bibinfo{author}{\bibfnamefont{L.}~\bibnamefont{Kn\"{o}ll}},
  \bibinfo{author}{\bibfnamefont{D.-G.} \bibnamefont{Welsch}},
  \bibinfo{author}{\bibfnamefont{S.}~\bibnamefont{Scheel}},
\bibnamefont{and}
  \bibinfo{author}{\bibfnamefont{J.}~\bibnamefont{K\"{a}stel}},
  \bibinfo{journal}{Phys. {R}ev. {A}} \textbf{\bibinfo{volume}{68}},
  \bibinfo{pages}{043816} (\bibinfo{year}{2003}).

\bibitem{0489} Ho Trung Dung, S.~Y. Buhmann, and D.-G. Welsch, Phys.
Rev. A \textbf{74}, 023803 (2006).

\bibitem{0739} A. Sambale, S. Y. Buhmann, D. -G. Welsch, and M. S.
Toma\v{s}, Phys. Rev. A \textbf{75}, 042109 (2007).

\bibitem{Ho01}
Ho Trung Dung, L.~Kn\"{o}ll, and D.-G. Welsch, Phys. Rev. A
\textbf{64}, 013804 (2001).

\bibitem{Ho00}
Ho Trung Dung, L.~Kn\"{o}ll, and D.-G. Welsch, Phys. Rev. A
\textbf{62}, 053804 (2000).

\bibitem[{\citenamefont{K\"{a}stel and
Fleischhauer}(2005)}]{0737}
\bibinfo{author}{\bibfnamefont{J.}~\bibnamefont{K\"{a}stel}},
  \bibinfo{author}{\bibfnamefont{M.}~\bibnamefont{Fleischhauer}},
  \bibinfo{journal}{Phys. {R}ev. {A}} \textbf{\bibinfo{volume}{71}},
  \bibinfo{pages}{011804(R)} (\bibinfo{year}{2005}).

\bibitem{0828} A. Sambale, S. Y. Buhmann, D.-G. Welsch, and Ho Trung
 Dung, Phys. Rev. A \textbf{78}, 053828 (2008).

\bibitem{0041}
J.~M. Wylie and J.~E. Sipe, Phys. Rev. A \textbf{30}, 1185 (1984).

\bibitem{Henkel99}
C.~Henkel and M.~Wilkens, Europhys. Lett. \textbf{47}, 414 (1999).

\bibitem{Henkel99b}
C.~Henkel, S. P\"{o}tting, and M.~Wilkens, Appl. Phys. B \textbf{69},
379 (1999).

\bibitem{0191}
P.~K. Rekdal, S.~Scheel, P.~L. Knight, and E.~A. Hinds, Phys. Rev. A
\textbf{70}, 013811 (2004).

\bibitem{Fermani07}
R.~Fermani, S.~Scheel, and P.~L. Knight, Phys. Rev. A \textbf{75},
062905 (2007).

\bibitem[{\citenamefont{Buhmann and Welsch}(2006)}]{0696}
\bibinfo{author}{\bibfnamefont{S.~Y.} \bibnamefont{Buhmann}}
\bibnamefont{and}
  \bibinfo{author}{\bibfnamefont{D.-G.} \bibnamefont{Welsch}},
  \bibinfo{journal}{Prog. {Q}uantum {E}lectron.} \textbf{\bibinfo{volume}{31}},
  \bibinfo{pages}{51} (\bibinfo{year}{2006}).

\bibitem[{\citenamefont{Kubo}(1966)}]{0751}
\bibinfo{author}{\bibfnamefont{R.}~\bibnamefont{Kubo}}, \bibinfo{journal}{Rep.
  {P}rog. {P}hys.} \textbf{\bibinfo{volume}{29}}, \bibinfo{pages}{255}
  (\bibinfo{year}{1966}).

\bibitem[{\citenamefont{Kn\"{o}ll et~al.}(2001)\citenamefont{Kn\"{o}ll, Scheel,
  and Welsch}}]{0003}
\bibinfo{author}{\bibfnamefont{L.}~\bibnamefont{Kn\"{o}ll}},
  \bibinfo{author}{\bibfnamefont{S.}~\bibnamefont{Scheel}}, \bibnamefont{and}
  \bibinfo{author}{\bibfnamefont{D.-G.} \bibnamefont{Welsch}}, in
  \emph{\bibinfo{booktitle}{Coherence and Statistics of Photons and Atoms}},
  edited by \bibinfo{editor}{\bibfnamefont{J.}~\bibnamefont{Pe\v{r}ina}}
  (\bibinfo{publisher}{Wiley}, \bibinfo{address}{New York},
  \bibinfo{year}{2001}), p.~\bibinfo{pages}{1}.

\bibitem[{\citenamefont{Brown and Carrington}(2003)}]{0813}
\bibinfo{author}{\bibfnamefont{J.}~\bibnamefont{Brown}} \bibnamefont{and}
  \bibinfo{author}{\bibfnamefont{A.}~\bibnamefont{Carrington}},
  \emph{\bibinfo{title}{Rotational Spectroscopy of Diatomic
 Molecules}}
  (\bibinfo{publisher}{Cambridge University Press},
  \bibinfo{address}{Cambridge}, \bibinfo{year}{2003}).

\bibitem[{\citenamefont{Bellini et~al.}(1995)\citenamefont{Bellini, De~Natale,
  Inguscio, Varberg, and Brown}}]{0817}
\bibinfo{author}{\bibfnamefont{M.}~\bibnamefont{Bellini}},
  \bibinfo{author}{\bibfnamefont{P.}~\bibnamefont{De~Natale}},
  \bibinfo{author}{\bibfnamefont{M.}~\bibnamefont{Inguscio}},
  \bibinfo{author}{\bibfnamefont{T.~D.} \bibnamefont{Varberg}},
  \bibnamefont{and} \bibinfo{author}{\bibfnamefont{J.~M.} \bibnamefont{Brown}},
  \bibinfo{journal}{Phys. {R}ev. {A}} \textbf{\bibinfo{volume}{52}},
  \bibinfo{pages}{1954} (\bibinfo{year}{1995}).

\bibitem[{\citenamefont{Rothstein}(1969)}]{0818}
\bibinfo{author}{\bibfnamefont{E.}~\bibnamefont{Rothstein}},
  \bibinfo{journal}{J. {C}hem. {P}hys.} \textbf{\bibinfo{volume}{50}},
  \bibinfo{pages}{1899} (\bibinfo{year}{1969}).

\bibitem[{\citenamefont{Ram et~al.}(1999)\citenamefont{Ram, Bernath, and
  Hinkle}}]{0798}
\bibinfo{author}{\bibfnamefont{R.~S.} \bibnamefont{Ram}},
  \bibinfo{author}{\bibfnamefont{P.~F.} \bibnamefont{Bernath}},
  \bibnamefont{and} \bibinfo{author}{\bibfnamefont{K.~H.}
  \bibnamefont{Hinkle}}, \bibinfo{journal}{J. {C}hem. {P}hys.}
  \textbf{\bibinfo{volume}{110}}, \bibinfo{pages}{5557} (\bibinfo{year}{1999}).

\bibitem[{\citenamefont{Campbell et~al.}(2008)\citenamefont{Campbell,
  Groenenboom, Lu, Tsi\-kata, and Doyle}}]{0799}
\bibinfo{author}{\bibfnamefont{W.~C.} \bibnamefont{Campbell}},
  \bibinfo{author}{\bibfnamefont{G.~C.} \bibnamefont{Groenenboom}},
  \bibinfo{author}{\bibfnamefont{H.-I.} \bibnamefont{Lu}},
  \bibinfo{author}{\bibfnamefont{E.}~\bibnamefont{Tsi\-kata}},
  \bibnamefont{and} \bibinfo{author}{\bibfnamefont{J.~M.} \bibnamefont{Doyle}},
  \bibinfo{journal}{Phys. {R}ev. {L}ett.} \textbf{\bibinfo{volume}{100}},
  \bibinfo{pages}{083003} (\bibinfo{year}{2008}).

\bibitem[{\citenamefont{Dieke and Crosswhite}(1962)}]{0804}
\bibinfo{author}{\bibfnamefont{G.~H.} \bibnamefont{Dieke}} \bibnamefont{and}
  \bibinfo{author}{\bibfnamefont{H.~M.} \bibnamefont{Crosswhite}},
  \bibinfo{journal}{J. {Q}uant. {S}pectrosc. {R}adiat. {T}ransf.}
  \textbf{\bibinfo{volume}{2}}, \bibinfo{pages}{97} (\bibinfo{year}{1962}).

\bibitem[{\citenamefont{Maillard et~al.}(1976)\citenamefont{Maillard,
  Chauville, and Mantz}}]{0807}
\bibinfo{author}{\bibfnamefont{J.~P.} \bibnamefont{Maillard}},
  \bibinfo{author}{\bibfnamefont{J.}~\bibnamefont{Chauville}},
  \bibnamefont{and} \bibinfo{author}{\bibfnamefont{A.~W.} \bibnamefont{Mantz}},
  \bibinfo{journal}{J. {M}ol. {S}pectrosc.} \textbf{\bibinfo{volume}{63}},
  \bibinfo{pages}{120} (\bibinfo{year}{1976}).

\bibitem[{\citenamefont{Meerts and Dynamus}(1973{\natexlab{a}})}]{0808}
\bibinfo{author}{\bibfnamefont{W.~L.} \bibnamefont{Meerts}} \bibnamefont{and}
  \bibinfo{author}{\bibfnamefont{A.}~\bibnamefont{Dynamus}},
  \bibinfo{journal}{Chem. {P}hys. {L}ett.} \textbf{\bibinfo{volume}{23}},
  \bibinfo{pages}{45} (\bibinfo{year}{1973}{\natexlab{a}}).

\bibitem[{\citenamefont{Nelson~Jr. et~al.}(1990)\citenamefont{Nelson~Jr.,
  Schiffmann, Nesbitt, Orlando, and Burkholder}}]{0809}
\bibinfo{author}{\bibfnamefont{D.~D.} \bibnamefont{Nelson~Jr.}},
  \bibinfo{author}{\bibfnamefont{A.}~\bibnamefont{Schiffmann}},
  \bibinfo{author}{\bibfnamefont{D.~J.} \bibnamefont{Nesbitt}},
  \bibinfo{author}{\bibfnamefont{J.~J.} \bibnamefont{Orlando}},
  \bibnamefont{and} \bibinfo{author}{\bibfnamefont{J.~B.}
  \bibnamefont{Burkholder}}, \bibinfo{journal}{J. {C}hem. {P}hys.}
  \textbf{\bibinfo{volume}{93}}, \bibinfo{pages}{7003} (\bibinfo{year}{1990}).

\bibitem[{\citenamefont{Meerts and Dynamus}(1973{\natexlab{b}})}]{0805}
\bibinfo{author}{\bibfnamefont{W.~L.} \bibnamefont{Meerts}} \bibnamefont{and}
  \bibinfo{author}{\bibfnamefont{A.}~\bibnamefont{Dynamus}},
  \bibinfo{journal}{Astrophys. {J}.} \textbf{\bibinfo{volume}{180}},
  \bibinfo{pages}{L93} (\bibinfo{year}{1973}{\natexlab{b}}).

\bibitem[{\citenamefont{Kaledin et~al.}(1999)\citenamefont{Kaledin, Bloch,
  McCarthy, and Field}}]{0819}
\bibinfo{author}{\bibfnamefont{L.~A.} \bibnamefont{Kaledin}},
  \bibinfo{author}{\bibfnamefont{J.~C.} \bibnamefont{Bloch}},
  \bibinfo{author}{\bibfnamefont{M.~C.} \bibnamefont{McCarthy}},
  \bibnamefont{and} \bibinfo{author}{\bibfnamefont{R.~W.} \bibnamefont{Field}},
  \bibinfo{journal}{J. {M}ol. {S}pectrosc.} \textbf{\bibinfo{volume}{197}},
  \bibinfo{pages}{289} (\bibinfo{year}{1999}).

\bibitem[{\citenamefont{Childs et~al.}(1984)\citenamefont{Childs, Goodman,
  Nielsen, and Pfeufer}}]{0820}
\bibinfo{author}{\bibfnamefont{W.~J.} \bibnamefont{Childs}},
  \bibinfo{author}{\bibfnamefont{L.~S.} \bibnamefont{Goodman}},
  \bibinfo{author}{\bibfnamefont{U.}~\bibnamefont{Nielsen}}, \bibnamefont{and}
  \bibinfo{author}{\bibfnamefont{V.}~\bibnamefont{Pfeufer}},
  \bibinfo{journal}{J. {C}hem. {P}hys.} \textbf{\bibinfo{volume}{80}},
  \bibinfo{pages}{2283} (\bibinfo{year}{1984}).

\bibitem[{\citenamefont{Sauer et~al.}(1996)\citenamefont{Sauer, Wang, and
  Hinds}}]{0821}
\bibinfo{author}{\bibfnamefont{B.~E.} \bibnamefont{Sauer}},
  \bibinfo{author}{\bibfnamefont{J.}~\bibnamefont{Wang}}, \bibnamefont{and}
  \bibinfo{author}{\bibfnamefont{E.~A.} \bibnamefont{Hinds}},
  \bibinfo{journal}{J. {C}hem. {P}hys.} \textbf{\bibinfo{volume}{105}},
  \bibinfo{pages}{7412} (\bibinfo{year}{1996}).

\bibitem[{\citenamefont{Barrow and Chojnicki}(1975)}]{0791}
\bibinfo{author}{\bibfnamefont{R.~F.} \bibnamefont{Barrow}} \bibnamefont{and}
  \bibinfo{author}{\bibfnamefont{A.~H.} \bibnamefont{Chojnicki}},
  \bibinfo{journal}{J. {C}hem. {S}oc. {F}araday {T}rans. 2}
  \textbf{\bibinfo{volume}{71}}, \bibinfo{pages}{728} (\bibinfo{year}{1975}).

\bibitem[{\citenamefont{Dolg et~al.}(1992)\citenamefont{Dolg, Stoll, and
  Preuss}}]{0790}
\bibinfo{author}{\bibfnamefont{M.}~\bibnamefont{Dolg}},
  \bibinfo{author}{\bibfnamefont{H.}~\bibnamefont{Stoll}}, \bibnamefont{and}
  \bibinfo{author}{\bibfnamefont{H.}~\bibnamefont{Preuss}},
  \bibinfo{journal}{Chem. {P}hys.} \textbf{\bibinfo{volume}{165}},
  \bibinfo{pages}{21} (\bibinfo{year}{1992}).

\bibitem[{\citenamefont{Korek et~al.}(2000{\natexlab{a}})\citenamefont{Korek,
  Allouche, Kobeissi, Chaalan, Dagher, Fakherddin, and
  Aubert-Fr\'{e}con}}]{0794}
\bibinfo{author}{\bibfnamefont{M.}~\bibnamefont{Korek}},
  \bibinfo{author}{\bibfnamefont{A.~R.} \bibnamefont{Allouche}},
  \bibinfo{author}{\bibfnamefont{M.}~\bibnamefont{Kobeissi}},
  \bibinfo{author}{\bibfnamefont{A.}~\bibnamefont{Chaalan}},
  \bibinfo{author}{\bibfnamefont{M.}~\bibnamefont{Dagher}},
  \bibinfo{author}{\bibfnamefont{K.}~\bibnamefont{Fakherddin}},
  \bibnamefont{and}
  \bibinfo{author}{\bibfnamefont{M.}~\bibnamefont{Aubert-Fr\'{e}con}},
  \bibinfo{journal}{Chem. {P}hys.} \textbf{\bibinfo{volume}{256}},
  \bibinfo{pages}{1} (\bibinfo{year}{2000}{\natexlab{a}}).

\bibitem[{\citenamefont{Tarnovsky et~al.}(1993)\citenamefont{Tarnovsky,
  Bunimovicz, Vu\v{s}kovi\'{c}, Stumpf, and Bederson}}]{0793}
\bibinfo{author}{\bibfnamefont{V.}~\bibnamefont{Tarnovsky}},
  \bibinfo{author}{\bibfnamefont{M.}~\bibnamefont{Bunimovicz}},
  \bibinfo{author}{\bibfnamefont{L.}~\bibnamefont{Vu\v{s}kovi\'{c}}},
  \bibinfo{author}{\bibfnamefont{B.}~\bibnamefont{Stumpf}}, \bibnamefont{and}
  \bibinfo{author}{\bibfnamefont{B.}~\bibnamefont{Bederson}},
  \bibinfo{journal}{J. {C}hem. {P}hys.} \textbf{\bibinfo{volume}{98}},
  \bibinfo{pages}{3894} (\bibinfo{year}{1993}).

\bibitem[{\citenamefont{Ross et~al.}(1990)\citenamefont{Ross, Effantin, Crozet,
  and Boursey}}]{0796}
\bibinfo{author}{\bibfnamefont{A.~J.} \bibnamefont{Ross}},
  \bibinfo{author}{\bibfnamefont{C.}~\bibnamefont{Effantin}},
  \bibinfo{author}{\bibfnamefont{P.}~\bibnamefont{Crozet}}, \bibnamefont{and}
  \bibinfo{author}{\bibfnamefont{E.}~\bibnamefont{Boursey}},
  \bibinfo{journal}{J. {P}hys. {B}: {A}t. {M}ol. {O}pt. {P}hys.}
  \textbf{\bibinfo{volume}{23}}, \bibinfo{pages}{L247} (\bibinfo{year}{1990}).

\bibitem[{\citenamefont{Korek et~al.}(2000{\natexlab{b}})\citenamefont{Korek,
  Allouche, Fakhreddine, and Chaa\-lan}}]{0795}
\bibinfo{author}{\bibfnamefont{M.}~\bibnamefont{Korek}},
  \bibinfo{author}{\bibfnamefont{A.~R.} \bibnamefont{Allouche}},
  \bibinfo{author}{\bibfnamefont{K.}~\bibnamefont{Fakhreddine}},
  \bibnamefont{and}
  \bibinfo{author}{\bibfnamefont{A.}~\bibnamefont{Chaa\-lan}},
  \bibinfo{journal}{Can. {J}. {P}hys.} \textbf{\bibinfo{volume}{78}},
  \bibinfo{pages}{977} (\bibinfo{year}{2000}{\natexlab{b}}).

\bibitem[{\citenamefont{Fellows et~al.}(1999)\citenamefont{Fellows, Gutteres,
  Campos, Verg\`{e}s, and Amiot}}]{0797}
\bibinfo{author}{\bibfnamefont{C.~E.} \bibnamefont{Fellows}},
  \bibinfo{author}{\bibfnamefont{R.~F.} \bibnamefont{Gutteres}},
  \bibinfo{author}{\bibfnamefont{A.~P.~C.} \bibnamefont{Campos}},
  \bibinfo{author}{\bibfnamefont{J.}~\bibnamefont{Verg\`{e}s}},
  \bibnamefont{and} \bibinfo{author}{\bibfnamefont{C.}~\bibnamefont{Amiot}},
  \bibinfo{journal}{J. {M}ol. {S}pectrosc.} \textbf{\bibinfo{volume}{197}},
  \bibinfo{pages}{19} (\bibinfo{year}{1999}).

\bibitem{endnote}
In comparing the results of Ref.~\cite{0772} with the free space limit
of our results, there are several comments worth making. (i)
\textit{Constants}. Because we have used measured values where
possible, there are small discrepancies between some of our value of
$\mu_\mathrm{e}$ and those used in Ref.~\cite{0772}. For KRb the
difference is large. These affect the rotational but not the
vibrational heating rates. Our values of $\mu'_\mathrm{e}$ are the
calculated ones presented in Tab.~I of Ref.~\cite{0772}. Note that the
value given there for LiRb should read 0.34\,(D$a_0^{-1}$), not
0.14\,(D$a_0^{-1}$). (ii) \textit{Heating}. We take room temperature
as 293\,K, rather than 300\,K, resulting in small differences
throughout. Table~II of Ref.~\cite{0772} gives lifetimes that are too
long for NaCs and KCs because they neglect rotational excitation,
which is the dominant effect. Our rotational heating rate for LiH
differs from the rate given in Tab.~II of Ref.~\cite{0772} because our
initial state is $J=0$, whereas theirs is $J=1$.

\bibitem[{\citenamefont{Mills et~al.}(1993)\citenamefont{Mills, Cvita\v{s},
  Homann, Kallay, and Kuchitsu}}]{0812}
\bibinfo{author}{\bibfnamefont{I.}~\bibnamefont{Mills}},
  \bibinfo{author}{\bibfnamefont{T.}~\bibnamefont{Cvita\v{s}}},
  \bibinfo{author}{\bibfnamefont{K.}~\bibnamefont{Homann}},
  \bibinfo{author}{\bibfnamefont{N.}~\bibnamefont{Kallay}}, \bibnamefont{and}
  \bibinfo{author}{\bibfnamefont{K.}~\bibnamefont{Kuchitsu}},
  \emph{\bibinfo{title}{Quantities, Units and Symbols in Physical Chemistry}}
  (\bibinfo{publisher}{Blackwell Scientific Publications},
  \bibinfo{address}{Oxford}, \bibinfo{year}{1993}).

\bibitem[{\citenamefont{Osterbrock et~al.}(1998)\citenamefont{Osterbrock,
  Fulbright, Cosby, and Barlow}}]{0811}
\bibinfo{author}{\bibfnamefont{D.~E.} \bibnamefont{Osterbrock}},
  \bibinfo{author}{\bibfnamefont{J.~P.} \bibnamefont{Fulbright}},
  \bibinfo{author}{\bibfnamefont{P.~C.} \bibnamefont{Cosby}}, \bibnamefont{and}
  \bibinfo{author}{\bibfnamefont{T.~A.} \bibnamefont{Barlow}},
  \bibinfo{journal}{Publ. {A}stron. {S}oc. {P}ac.}
  \textbf{\bibinfo{volume}{110}}, \bibinfo{pages}{1499} (\bibinfo{year}{1998}).

\bibitem[{\citenamefont{Coxon}(1975)}]{0806}
\bibinfo{author}{\bibfnamefont{J.~A.} \bibnamefont{Coxon}},
  \bibinfo{journal}{J. {M}ol. {S}pectrosc.} \textbf{\bibinfo{volume}{58}},
  \bibinfo{pages}{1} (\bibinfo{year}{1975}).

\bibitem[{\citenamefont{Palik}(1991)}]{0814}
\bibinfo{editor}{\bibfnamefont{E.~D.} \bibnamefont{Palik}}, ed.,
  \emph{\bibinfo{title}{Handbook of Optical Constants of Solids {I}{I}}}
  (\bibinfo{publisher}{Academic Press}, \bibinfo{address}{New York},
  \bibinfo{year}{1991}).

\bibitem[{\citenamefont{Franzen}(2008)}]{0815}
\bibinfo{author}{\bibfnamefont{S.}~\bibnamefont{Franzen}},
  \bibinfo{howpublished}{private communication} (\bibinfo{year}{2008}).

\bibitem{Grignon(1)2003} R.~Grignon, M.~N. Asfar, Y.~Wang and S.~Butt,
Proceedings of the 20th IEEE Instrumentation and Measurement
Technology Conference, Vol. 1, p.~865, IEEE (2003)

\bibitem[{\citenamefont{Pendry et~al.}(1999)\citenamefont{Pendry, Holden,
  Robbins, and Stewart}}]{0477}
\bibinfo{author}{\bibfnamefont{J.~B.} \bibnamefont{Pendry}},
  \bibinfo{author}{\bibfnamefont{A.~J.} \bibnamefont{Holden}},
  \bibinfo{author}{\bibfnamefont{D.~J.} \bibnamefont{Robbins}},
  \bibnamefont{and} \bibinfo{author}{\bibfnamefont{W.~J.}
  \bibnamefont{Stewart}}, \bibinfo{journal}{I{EEE} {T}rans. {M}icrow. {T}heory
  {T}ech.} \textbf{\bibinfo{volume}{47}}, \bibinfo{pages}{2075}
  (\bibinfo{year}{1999}).

\bibitem[{\citenamefont{Smith et~al.}(2000)\citenamefont{Smith, Padilla, Vier,
  Nemat-Nasser, and Schultz}}]{0479}
\bibinfo{author}{\bibfnamefont{D.~R.} \bibnamefont{Smith}},
  \bibinfo{author}{\bibfnamefont{W.~J.} \bibnamefont{Padilla}},
  \bibinfo{author}{\bibfnamefont{D.~C.} \bibnamefont{Vier}},
  \bibinfo{author}{\bibfnamefont{S.~C.} \bibnamefont{Nemat-Nasser}},
  \bibnamefont{and} \bibinfo{author}{\bibfnamefont{S.}~\bibnamefont{Schultz}},
  \bibinfo{journal}{Phys. {R}ev. {L}ett.} \textbf{\bibinfo{volume}{84}},
  \bibinfo{pages}{4184} (\bibinfo{year}{2000}).

\bibitem[{\citenamefont{Veselago}(1968)}]{0476}
\bibinfo{author}{\bibfnamefont{V.~G.} \bibnamefont{Veselago}},
  \bibinfo{journal}{Sov. {P}hys. {U}spekhi} \textbf{\bibinfo{volume}{10}},
  \bibinfo{pages}{509} (\bibinfo{year}{1968}).

\bibitem{Lenac}
Z.~Lenac and M.~S. Toma\v{s}, Surf. Sci. \textbf{215}, 299
(1989); R.~M. Amos and W.~L. Barnes, Phys. Rev. B \textbf{55}, 7249 (1997).

\bibitem{Naftaly07}
M.~Naftaly and R.~E. Miles, Proc. IEEE \textbf{95}, 1658 (2007).


\end{thebibliography}
\end{document}